\documentclass[a4paper,10pt,twocolumn,accepted=2022-05-28]{quantumarticle}

\pdfoutput=1
\usepackage{amsmath}
\usepackage{graphicx}
\usepackage{wasysym}
\usepackage{amsfonts}
\usepackage{amssymb}
\usepackage{bm}
\usepackage{enumerate}
\usepackage{color}
\usepackage[colorlinks=true,
            linkcolor=magenta,
            urlcolor=blue,
            citecolor=blue]{hyperref}

\usepackage[numbers,sort&compress]{natbib}
\newcommand{\beq}{\begin{equation}}
\newcommand{\eeq}{\end{equation}}
\newcommand{\bea}{\begin{aligned}}
\newcommand{\eea}{\end{aligned}}
\newcommand{\bes}{\begin{split}}
\newcommand{\ees}{\end{split}}

\newcommand{\RNum}[1]{\uppercase\expandafter{\romannumeral #1\relax}}

\newcommand{\la}{\langle}
\newcommand{\ra}{\rangle}

\mathchardef\nss="711B

\def\ket#1{{\left|#1\right\rangle}}
\def\bra#1{{\left\langle #1 \right|}}

\def\nss{\mathcal{S}}

\def\be{\begin{eqnarray}}
\def\ee{\end{eqnarray}}

\newlength{\myL}

\begin{document}

\title{Integrability of the $\nu=4/3$ fractional quantum Hall edge states}
\author{Yichen Hu}
\affiliation{Department of Physics, Princeton University, Princeton, New Jersey 08544, USA}
\affiliation{The Rudolf Peierls Centre for Theoretical Physics, University of Oxford, Oxford OX1 3PU, UK}
\author{Biao Lian}
\affiliation{Department of Physics, Princeton University, Princeton, New Jersey 08544, USA}

\begin{abstract}
We investigate the homogeneous chiral edge theory of the filling $\nu=4/3$ fractional quantum Hall state, which is parameterized by a Luttinger liquid velocity matrix and an electron tunneling amplitude (ignoring irrelevant terms). We identify two solvable cases: one case where the theory gives two free chiral boson modes, and the other case where the theory yields one free charge $\frac{2e}{\sqrt{3}}$ chiral fermion and two free chiral Bogoliubov (Majorana) fermions. For generic parameters, the energy spectrum from our exact diagonalization shows Poisson level spacing statistics (LSS) in each conserved charge and momentum sector, indicating the existence of hidden conserved quantities and the possibility that the generic edge theory of the $\nu=4/3$ fractional quantum Hall state is integrable. We further show that a global symmetry preserving irrelevant nonlinear kinetic term will lead to the transition of LSS from Poisson to Wigner-Dyson at high energies. This further supports the possibility that the model without irrelevant terms is integrable.
\end{abstract}
\maketitle

\section{Introduction}

The integrability of quantum systems is known to significantly affect their quantum coherence and quantum dynamics. An important class of integrable quantum models is the chiral Luttinger liquid \cite{Dzyaloshinski1996,Haldane_1981,Haldane1981,Tomonaga1950,Luttinger1963,Wen1,Wen2},
which are free boson theories describing the low-energy chiral edge states of many fractional quantum Hall (FQH) states \cite{Wen1,Wen2}. The quantum coherence of integrable chiral edge states makes it possible to observe their interference \cite{Willett2009,Zhao2020,Lian2016,Roulleau2008}, 
which for instance enables the detection of anyon braiding in the Fabry-P\'{e}rot interferometers of the filling $\nu=1/3$ FQH state \cite{Laughlin1983,Nakamura2020,Carrega2021,McClure2012,Ofek2010,Halperin2011}. 
On the other hand, interactions between multiple edge modes may lead to mode reconstructions \cite{Kane1994,Levin2007,Lee2007,Lian2018,Chamon1994,Wan2002,Wan2003,Sabo2017,Cano2014}, robust quantum chaos at low energies \cite{lian2019,hu2021chiral}, and quantum scars \cite{Frank2022,martin2021scar}, which are crucial for understanding their thermal equilibration in thermal transports \cite{Banerjee2018,Feldman2018,Simon2018,Ma2019}.

An intriguing question is whether low-energy chiral edge states can be integrable without being free boson/fermion theories. In this letter, we investigate the chiral edge theory of the filling $\nu=4/3$ FQH state \cite{Boebinger1985,Lai2004,Ismail1995,Maiti2021,Pudalov1984,Dean2011} as a possible example of such, which is parameterized by a Luttinger liquid velocity matrix and a dimensionless electron tunneling amplitude, with irrelevant terms ignored. We identify analytically two solvable special cases: one case where the theory gives two free chiral bosons, and another case where the theory consists of a charge $\frac{2e}{\sqrt{3}}$ free chiral fermion and two charge neutral free chiral Bogoliubov (Majorana) fermions. For generic parameters, we employ exact diagonalization (ED) to numerically calculate the energy spectrum. Interestingly, our results show that the many-body energy spectrum of each conserved global charge and momentum sector obeys a Poisson level spacing statistics. This indicates that there are hidden local conserved quantities, and possibly the $4/3$ FQH edge model is integrable while not being a free theory. We further test the effect of an irrelevant nonlinear kinetic term that preserves all global symmetries. For large nonlinearity, LSS shifts from Poisson  towards Wigner-Dyson, indicating the breaking of the potential integrability of the model by irrelevant terms at high energies. Lastly, we identify a family of FQH edge theories which share similar physics.

\section{The edge model}
The filling $\nu=4/3$ FQH state can be viewed as a filling $\nu=1/3$ Laughlin FQH state plus a filling $\nu=1$ integer QH (IQH) state in the bulk. Accordingly, its edge theory consists of two forward-moving chiral bosons modes: $\phi_1$ and $\phi_2$ from the $1/3$ Laughlin state and the IQH state, respectively, which have a free Lagrangian density
\begin{equation} \label{Lag0}
\mathcal{L}_0=-\frac{1}{4\pi}\sum_{i,j=1}^2\partial_x \phi_i(K_{ij}\partial_t+V_{ij}\partial_x)\phi_j\ ,
\end{equation}
where $K_{ij}=\text{diag}(3,1)$ is the $K$ matrix with integer entries, and $V_{ij}$ is the real symmetric velocity matrix. The physical edge charge excitations are given by vertex operators $e^{i(l_1\phi_1+l_2\phi_2)}$, where $\bm{l}=(l_1,l_2)^T$ is an integer vector, and their electrical charges are $Q_{\bm{l}}=\bm{l}^T K^{-1}\bm{q}$, with $\bm{q}=(1,1)^T$ being the charge vector associated with the bosons \cite{Wen1,Wen2}. In particular, $\chi=e^{i\phi_1}$ is the annihilation operator of the charge $e/3$ anyon of the Laughlin state with exchange statistical phase $\pi/3$. The operators annihilating an electron on the edge are $\psi'=e^{3i\phi_1}$ and $\psi=e^{i\phi_2}$.

Electron tunneling between the edges of the $1/3$ Laughlin state and the IQH state generically exists. Here we assume the edge is homogeneous, and assume the two modes $\phi_1,\phi_2$ have a momentum difference $p$, then the leading electron tunneling term takes the form
\begin{equation}\label{Lag1}
\mathcal{L}_1=-J(x) e^{3i\phi_1-i\phi_2}-h.c.,\qquad J(x)=Je^{ipx}\ ,
\end{equation}
where $J$ is constant. The chiral operator $e^{3i\phi_1-i\phi_2}$ has an exact scaling dimension $2$ as fixed by its conformal spin, making the coupling $J$ dimensionless and non-negligible at low energies. Note that a similar tunneling term occurs on the $2/3$ FQH edge \cite{Kane1994}, but there the term has a velocity-dependent scaling dimension because the edge is not fully chiral.

In realistic experimental systems such as the GaAs/AlGaAs 2D electron gas, since the cyclotron energy is much larger than the Zeeman energy, the monolayer 4/3 FQH state usually occupies two Landau levels with opposite spins (filling $1$ for spin up and filling $1/3$ for spin down). This would imply the electron tunneling term in Eq. ~(\ref{Lag1}) is spin flipping, which is possible in the presence of spin-orbital coupling (SOC). In GaAs/AlGaAs, the edge mode velocity parameters $V_{ij}$ in our model are typically given by the Coulomb interaction as of order $\sim 0.2 e^2/\epsilon\approx 20$ meV $\cdot$ nm \cite{Yang2009}, where $\epsilon\approx 13$ is the dielectric constant. The spin-flipping tunneling $J$ term is of the order of the spin-flipping SOCs in GaAs/AlGaAs, for instance, the Dresselhaus SOC parameter around $\sim 4$ meV$\cdot$ nm, and the Rashba SOC parameter within a similar range depending on the displacement field of the quantum well \cite{Rozhansky2016}. The tunneling parameter $J$ is thus not far off from the velocity parameters $V_{ij}$.  We do note that the spatial separation of different FQH edge states may make $J$ smaller. If instead the 4/3 FQH state is realized in a bilayer structure and occupies two Landau levels with the same spin (with fillings $1$ and $1/3$, respectively), the electron tunneling $J$ term would not require spin flipping SOC. However, this also requires an asymmetric filling which is less common in bilayer FQH experiments, and the layer separation may further suppress the tunneling.

The above model Lagrangian $\mathcal{L}=\mathcal{L}_0+\mathcal{L}_1$ can be mapped into an interacting fermion model. We first transform into a new boson basis
\begin{equation}\label{eq:phinrho}
\phi_n=\frac{3\phi_1-\phi_2}{2}\ ,\quad \phi_{\rho}=\frac{\sqrt{3}(\phi_1+\phi_2)}{2}\ ,
\end{equation}
under which the $K$ matrix in Eq. (\ref{Lag0}) transforms into an identity matrix. This allows us to define two fermion annihilation fields $c_n=e^{i\phi_n}$ and $c_\rho=e^{i\phi_\rho}$ with scaling dimension $\frac{1}{2}$. The fermion $c_n$ is charge neutral, while $c_\rho$ carries an irrational electron charge $\frac{2e}{\sqrt{3}}$. The Lagrangian density then fermionizes into (App. ~\ref{App:A})
\begin{equation}
\begin{aligned}
\label{eq:LaF}
&\mathcal{L}=ic^\dagger_n(\partial_t+v_n\partial_x)c_n+ic_\rho^\dagger(\partial_t+v_\rho \partial_x)c_\rho\\
&+\frac{J e^{ipx}}{2\pi}ic_n \partial_x c_n+\frac{J^* e^{-ipx}}{2\pi}ic_n^\dag \partial_x c_n^\dag-2\pi v_{n\rho} c^\dagger_n c_n c^\dagger_\rho c_\rho\ ,
\end{aligned}
\end{equation} 
where $v_\rho=\frac{V_{11}+6V_{12}+9V_{22}}{12}$, $v_n=\frac{V_{11}-2V_{12}+V_{22}}{4}$ and $v_{n\rho}=\frac{V_{11}+2V_{12}-3V_{22}}{4\sqrt{3}}$, and we have used the bosonization mappings between fermions and bosons $c^\dag_\alpha c_\alpha =\frac{\partial_x\phi_\alpha}{2\pi}$, $-ic^\dag_\alpha\partial_x c_\alpha =\frac{(\partial_x\phi_\alpha)^2}{4\pi}$, and $-ic_\alpha\partial_x c_\alpha =2\pi e^{2i\phi_\alpha}$ ($\alpha=n,\rho$) \cite{lian2019,hu2021chiral}. In the rest of paper, we study the eigen-spectrum of the model in Eq. (\ref{eq:LaF}), as well as calculate the (time non-ordered) zero-temperature fermion two-point functions $G_{n/\rho}(t,x)=\langle c_{n/\rho}(t,x)c_{n/\rho}^\dag (0,0)\rangle$, and the two-point functions $G_\chi (t,x)=\langle \chi(t,x)\chi^\dag(0,0) \rangle$ of the charge $e/3$ anyon operator $\chi=e^{i\phi_1}$ and $G_\psi (t,x)=\langle \psi(t,x)\psi^\dag(0,0) \rangle$ of the electron operator $\psi=e^{i\phi_2}$. We first focus on two solvable free points, and then turn to a generic exact diagonalization (ED) study.

\section{The exactly solvable free limits}
\subsection{The $J=0$ case}
When the tunneling term $\mathcal{L}_1$ is absent, namely $J=0$, the Lagrangian $\mathcal{L}=\mathcal{L}_0$ in Eq. (\ref{Lag0}) yields two free chiral boson modes $\phi_+'=\phi_\rho\cos\frac{\zeta}{2}+\phi_n\sin\frac{\zeta}{2}$ and $\phi_-'=-\phi_\rho\sin\frac{\zeta}{2}+\phi_n\cos\frac{\zeta}{2}$ with $\zeta=\arctan\frac{2v_{n\rho}}{v_\rho-v_n}$. Both modes have linear dispersions $\omega_\pm(k)=v_\pm'k$, with the velocities $v_\pm'=\frac{v_n+ v_\rho\pm \sqrt{4v_{n\rho}^2+(v_n-v_\rho)^2}}{2}$, respectively. The model in Eq. (\ref{eq:LaF}) at $J=0$ is therefore integrable. The correlations of the boson fields are given by $\la \phi'_\eta (t,x) \phi'_{\eta'}(0,0) \ra=-\delta_{\eta\eta'}\log{[2\pi i(v'_\eta t-x-i0^+)]}$, where $\eta,\eta'=\pm$ and $0^+$ an infinitesimal positive number. Accordingly, the two-point functions of different particle operators can be obtained from the Wick's theorem as
\begin{equation}
G_\alpha (t,x)=\prod_{\eta=\pm}\frac{1}{[2\pi i(v'_\eta t-x-i0^+)]^{\sigma_{\alpha}^\eta}}\ ,
\end{equation}
where $\alpha=n,\rho,\chi,\psi$ 
denotes the particle species we defined earlier. The corresponding exponents are given by $\sigma_n^\pm=\frac{1\mp\cos\zeta}{2}$, $\sigma_\rho^\pm=\frac{1\pm\cos\zeta}{2}$, $\sigma_\chi^\pm=\frac{2\pm(\sqrt{3}\sin{\zeta}-\cos{\zeta})}{12}$ and $\sigma_\psi^\pm=\frac{2\mp(\sqrt{3}\sin{\zeta}-\cos{\zeta})}{4}$ (App. \ref{App:B}). For fermions $\alpha=n,\rho,\psi$, we have $\sigma_\alpha^+ +\sigma_\alpha^-=1$, while for anyon $\chi$ we have $\sigma_\chi^+ +\sigma_\chi^-=\frac{1}{3}$, which yield the expected exchange statistical angles $\pi$ of the fermions and $\pi/3$ of anyon $\chi$, respectively.

\begin{figure}[tbp]
\centering\includegraphics[width=0.5\textwidth]{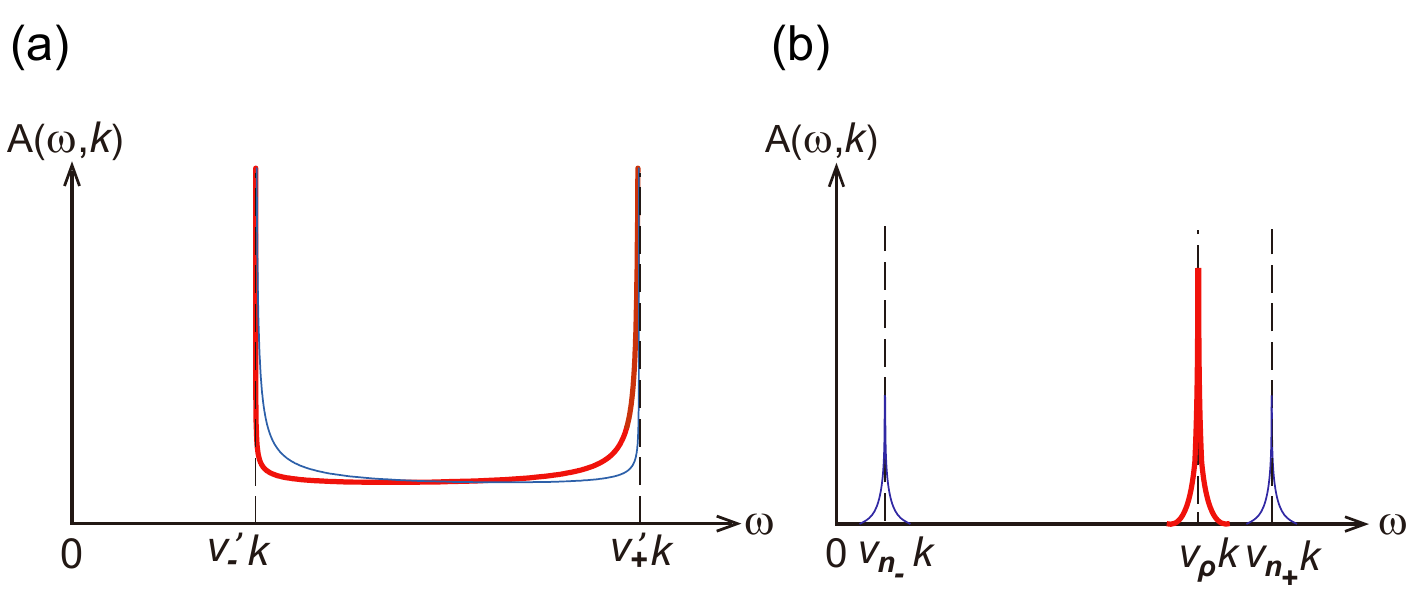}\caption{(a) Spectral weight $A_{\alpha=n,\rho}(\omega,k)$ at $J=0$, $k=7.5$, $v_\rho=1.5$, $v_n=1$ and $v_{n\rho}=0.8$. (b) Spectral weight $A_{\alpha=n,\rho}(\omega,k)$, at $J=0.8 \pi$, $p=0$, $k=7.5$, $v_\rho=1.5$, $v_n=1$ and $v_{n\rho}=0$. Red thick lines (Blue thin lines) stand for the $\rho$ ($n$) sector.}
\label{fig:sw}
\end{figure}

The two-point functions for fermions $\alpha=n,\rho,\psi$ allow us to define the retarded Green's function $G_{\alpha,\text{ret }}(t,x)=\Theta(t)[G_{\alpha}(t-i0^+,x)-G_{\alpha}(t+i0^+,x)]$. From its Fourier transform, we can define the spectral weight $A_\alpha(\omega,k)=2\text{Im}G_{\alpha,\text{ret}}(\omega,k)$, which vanishes when $\frac{\omega}{k}>v_+'$ and $\frac{\omega}{k}<v_-'$. An example of spectral weights of $n$ and $\rho$ fermions (thin and thick lines, respectively) at fixed $k>0$ in this case is plotted in Fig. \ref{fig:sw}(a).

\subsection{The $v_{n\rho}=0$ case} At the special point $v_{n\rho}=0$, the model in the form of Eq. (\ref{eq:LaF}) becomes a free fermion model, with fermions $n$ and $\rho$ decoupled. The fermion $c_\rho$ has a linear dispersion $\omega_\rho(k)=v_\rho k$. By changing the fermion $c_n$ to a momentum shifted basis $c_n'(x)=e^{ipx/2}c_n(x)$, which eliminates the $x$ dependence of tunneling term $J$, it splits into two chiral Bogoliubov fermionic modes $b_{n}(k)$ and $b_{n}^\dag(-k)$ in the Nambu basis, where the fermion $b_n(k)$ has a dispersion $\omega_{b_n}(k)=v_n k+\frac{1}{2\pi} \sqrt{(2Jk)^2+(\pi v_np)^2}$. 
The model in Eq. (\ref{eq:LaF}) in this case is thus integrable. 

We derive below the two-point functions of different particles in this case for \emph{momentum difference $p=0$}. The generic $p\neq0$ case is discussed in Appendix \ref{App:C}. 1. With $p=0$, the two Bogoliubov modes can be recombined into two Majorana modes $\gamma_{n_+}=\frac{c_n+c_n^\dag}{\sqrt{2}}$ and $\gamma_{n_-}=\frac{c_n-c_n^\dag}{\sqrt{2}i}$ with linear dispersions $\omega_{n_\pm}(k)=v_{n_\pm}k$, where velocities $v_{n_\pm}=v_n\pm \frac{J}{\pi}$. The two-point functions of the $n$ and $\rho$ fermions are simply free propagators:
\begin{equation}\label{eq:freef-GnGr}
\bea
&G_n(t,x)=\frac{1}{4\pi i}\displaystyle\sum_{s=\pm}\frac{1}{v_{n_s}t-x-i0^+}\ ,\\
&G_\rho(t,x)=\frac{1}{2\pi i}\frac{1}{v_{\rho}t-x-i0^+}\ .
\eea
\end{equation}
Their spectral weights are $A_{n}(\omega,k)=\pi(\delta(\omega-k v_{n_+})+\delta(\omega-k v_{n_-}))$ and $A_\rho(\omega,k)=2\pi\delta(\omega-k v_{\rho})$, respectively, as shown in Fig. \ref{fig:sw}(b).

In the bosonic representation, with $v_{n\rho}=0$, the boson $\phi_\rho$ is free and thus has a correlation $\langle \phi_\rho(t,x)\phi_\rho(0,0)\rangle=-\log{[2\pi i(v'_\eta t-x-i0^+)]}$. The boson $\phi_n$ is not free, although $c_n=e^{i\phi_n}$ splits into free Majorana fermions. It is thus easy to calculate the two-point function of the charge $2e/3$ anyon operator $\chi'=e^{2i\phi_1}=e^{i(\phi_n+\frac{\phi_\rho}{\sqrt{3}})}=c_n e^{i\frac{\phi_\rho}{\sqrt{3}}}$, which reads $G_{\chi'}(t,x)=G_n(t,x)e^{\frac{1}{3}\langle\phi_\rho(t,x)\phi_\rho(0,0)\rangle}=G_n(t,x)G_\rho(t,x)^{1/3}$. However, the calculation of two-point functions of the charge $e/3$ anyon operator $\chi=e^{i\phi_1}=e^{i(\frac{\phi_n}{2}+\frac{\phi_\rho}{2\sqrt{3}})}$ and the electron operator $\psi=e^{i\phi_2}=e^{i(\frac{\sqrt{3}\phi_\rho}{2}-\frac{\phi_n}{2})}$ are more complicated, due to the non-free nature of boson field $\phi_n$. We now show that they can be calculated by doubling the model. 

We define a double-copy of our model Lagrangian as $\mathcal{L}_{2c}=\mathcal{L}^{(+)}+\mathcal{L}^{(-)}$, where $\mathcal{L}^{(\eta)}$ is the Lagrangian $\mathcal{L}=\mathcal{L}_0+\mathcal{L}_1$ defined in Eqs. (\ref{Lag0}) and (\ref{Lag1}) with boson fields $\phi_1^{(\eta)}$ and $\phi_2^{(\eta)}$ ($\eta=\pm$). Accordingly, we perform a basis transformation $\phi_n^{(\eta)}=\frac{3\phi_1^{(\eta)}-\phi_2^{(\eta)}}{2}, \phi_\rho^{(\eta)}=\frac{\sqrt{3}(\phi_1^{(\eta)}+\phi_2^{(\eta)})}{2}$ for each copy, similar to Eq. (\ref{eq:phinrho}). Next, we transform to a boson basis $\widetilde{\phi}_j$ ($1\le j\le 4$) defined by $\phi_n^{(\eta)}=\frac{(\widetilde{\phi}_1-\widetilde{\phi}_2)+\eta(\widetilde{\phi}_3-\widetilde{\phi}_4)}{2}$ and $\phi_\rho^{(\eta)}=\frac{(\widetilde{\phi}_1+\widetilde{\phi}_2)+\eta(\widetilde{\phi}_3+\widetilde{\phi}_4)}{2}$, and define four complex fermion fields $f_j=e^{i\widetilde{\phi}_j}$. This is known as the SO(8) triality transformation \cite{shankar1981,shankar1983,Kane2020,Ryu2012,Maldacena1997,Kitaev2010,hu2021chiral}. 
By further transforming into a new fermion basis $d_{1,\pm}=\frac{f_1\pm i f_2}{\sqrt{2}}$, $d_{2,\pm}=\frac{f_3\pm i f_4}{\sqrt{2}}$, and then do a bosonization $d_{s,\eta}=e^{i\theta_{s,\eta}}$ to four boson fields $\theta_{s,\eta}$ ($s=1,2$, $\eta=\pm$), one arrives at a free boson model (see App. \ref{App:C} 2. b). The eigen-boson fields of this free boson model are $\widetilde{\theta}_{\eta \eta'}=(\theta_{1,+}+\eta\theta_{1,-}+\eta'\theta_{2,+}+\eta\eta'\theta_{2,-})$ where $\eta,\eta'=\pm$, under which the double-copy Lagrangian takes a diagonal form
\begin{equation}\label{eq:freef-to-freeb}
\bea
    \mathcal{L}_{2c}=-\frac{1}{4\pi}\sum_{\eta,\eta'=\pm} \partial_x \widetilde{\theta}_{\eta \eta'} (\partial_t+u_{\eta \eta'}\partial_x)\widetilde{\theta}_{\eta \eta'}\ .
\eea
\end{equation} 
The velocities $u_{\eta\eta'}$ are defined by $u_{+\pm}=v_\rho$, and  $u_{-\pm}=v_{n_\pm}$, with $v_\rho$ and $v_{n_\pm}$ in Eq. (\ref{eq:freef-GnGr}).

We can now calculate the two-point functions of the charge $e/3$ anyon $\chi=e^{i\phi_1}$ and the electron $\psi=e^{i\phi_2}$. Since the two copies we introduced are identical and decoupled, the two-point function of anyon $\chi$ can be rewritten as $G_{\chi}(t,x)=\sqrt{\langle \chi^{(+)}(t,x)\chi^{(-)}(t,x)\chi^{(+)\dag}(0,0)\chi^{(-)\dag}(0,0)\rangle}$, where $\chi^{(\pm)}=e^{i\phi_1^{(\pm)}}$ is the anyon operator in copy $\mathcal{L}^{(\pm)}$. Next, we note that the product operator  $\chi^{(+)}\chi^{(-)}=e^{i\frac{\phi_n^{(+)}+\phi_n^{(-)}}{2}+i\frac{\phi_\rho^{(+)}+\phi_\rho^{(-)}}{2\sqrt{3}}}$, and $\phi_\rho^{(\pm)}$ are free bosons and decoupled with $\phi_n^{(\pm)}$. Therefore, we only need to know the two-point function of operator $e^{i\frac{\phi_n^{(+)}+\phi_n^{(-)}}{2}}$, which can be derived from the two-point function of fermion $f_1=e^{i\widetilde{\phi}_1}=e^{i\frac{\phi_n^{(+)}+\phi_n^{(-)}+\phi_\rho^{(+)}+\phi_\rho^{(-)}}{2}}$.
This enables us to express the anyon two-point function as $G_\chi(t,x)=\sqrt{\langle f_1(t,x)f_1(0,0)\rangle G_\rho(t,x)^{-1/3}}$, with $G_\rho(t,x)$ given by Eq. (\ref{eq:freef-GnGr}). A similar transformation holds for the two-point function of electron $\psi$. The two-point function of fermion $f_1$ can be calculated by transforming into the fermion basis $d_{s,\eta}=e^{i\theta_{s,\eta}}$, which yields $\langle f_1(t,x)f_1(0,0)\rangle=\prod_{\eta,\eta'=\pm}e^{\frac{\langle\widetilde{\theta}_{\eta\eta'}(t,x)\widetilde{\theta}_{\eta\eta'}(0,0)\rangle}{4}}$ in terms of the free bosons $\widetilde{\theta}_{\eta\eta'}$ in Eq. (\ref{eq:freef-to-freeb}). We therefore find the two-point function of anyon $\chi$ and electron $\psi$ given by (App. \ref{App:C}. 2. b)
\begin{equation}
G_\alpha(t,x)=\displaystyle\prod_{\eta=\rho,n_\pm}\frac{1}{[2\pi i(v_\eta t-x-i0^+)]^{\sigma_\alpha^\eta}}\ ,
\end{equation} 
where $\alpha=\chi,\psi$, and the exponents $\sigma_\chi^\rho=\frac{1}{12}$, $\sigma_\psi^\rho=\frac{3}{4}$,  $\sigma_\chi^{n_\pm}=\sigma_\psi^{n_\pm}=\frac{1}{8}$. 


\section{The generic case}
For generic parameters $J\neq0$ and $v_{n\rho}\neq0$, the model in Eq. (\ref{eq:LaF}) is difficult to solve, so we turn to study the model numerically with ED. To simplify the problem, we first eliminate $e^{ipx}$ factor of the tunneling $J$ term by a unitary transformation, at the price of adding a chemical potential. This is done by shifting the boson fields $\phi_n$ and $\phi_\rho$ in Eq. (\ref{eq:phinrho}) into a new basis $\phi_n'=\phi_n+\frac{p}{2}x$ and $\phi_\rho'=\phi_\rho -\frac{v_n p}{2v_{n\rho}}x$, and define their corresponding fermion fields $c_n'=e^{i\phi_n'}=e^{ipx/2}c_n$ and $c_\rho'=e^{i\phi_\rho'}=e^{-i\frac{v_n p}{2v_{n\rho}}x}c_\rho$. The Lagrangian in Eq. (\ref{eq:LaF}) then becomes (App. \ref{App:D}. 1)
\begin{equation}\label{eq:LaF2}
\begin{split}
&\mathcal{L}=ic_n'^\dagger(\partial_t+v_n\partial_x)c_n'+ic_\rho'^\dagger(\partial_t+v_\rho \partial_x)c_\rho'\\
&-\mu_\rho c_\rho'^\dagger c_\rho'+\left(i\frac{J}{2\pi}c_n' \partial_x c_n'+h.c.\right)-2\pi v_{n\rho} c_n'^\dagger c_n' c_\rho'^\dagger c_\rho'\ ,
\end{split}
\end{equation} 
where the chemical potential $\mu_\rho=\left(\frac{v_nv_\rho}{v_{n\rho}}-v_{n\rho}\right)p$.  Since the fermion $c_n'$ is neutral and the fermion $c_\rho'$ carry charge $\frac{2e}{\sqrt{3}}$, the $\mu_\rho$ term is equivalent to a physical chemical potential $\frac{2}{\sqrt{3}}\mu_\rho$ on the edge of the $4/3$ FQH state.

The model in the form of Eq. (\ref{eq:LaF2}) has three apparent conserved global charges: (i) the total many-body momentum $K_{\text{tot}}$, (ii) the total charge of the $\rho$ fermion $N_\rho$, and (iii) the parity of the $n$ fermion $(-1)^{N_n}$, which are explicitly given by
\begin{equation}
\begin{split}
&K_{\text{tot}}=-i\int dx (c_n'^\dag \partial_x c_n'+c_\rho'^\dag \partial_x c_\rho')\ ,\\
&N_\rho=\int dx c_\rho'^\dag c_\rho'\ , \\ 
&(-1)^{N_n}=(-1)^{\int dx c_n'^\dag c_n'}\ .
\end{split}
\end{equation}
To perform ED calculations, we assume the 1D space is closed and has length $L$, and we take an anti-periodic boundary condition for both the $c_n'$ and $c_\rho'$ fermions (which is the boundary condition for zero bulk flux). Since the chirality allows only positive momentum excitations, the Hilbert space dimension of each total momentum $(K_{\text{tot}}$ is finite, making our ED numerical calculation possible (see App. \ref{App:D}. 2).  For convenience, we set the unit of length such that $L=2\pi$. This constraints the single-particle momentum of the fermions to half-odd integers, i.e., $k\in \mathbb{Z}+\frac{1}{2}$. As a result, the many-body momentum $K_{\text{tot}}$ is an integer (half-odd integer) if the total number of fermions is even (odd). Therefore, the three conserved charges $K_{\text{tot}}$, $N_\rho$ and $(-1)^{N_n}$ are not independent, and we have 
\begin{equation}
(-1)^{N_n}=(-1)^{2K_{\text{tot}}+N_\rho}\ . 
\end{equation}

The $N_\rho=0$ charge sector has an additional special symmetry. To see this, we define a particle-hole transformation $P$ obeying
\beq\label{eq-P-def}
\begin{aligned}
&P c_{\alpha} P^{-1}=c^\dag_{\alpha},\quad P c^\dag_{\alpha} P^{-1}=c_{\alpha},\quad (\alpha=\rho, n)\\
&P N_{\rho} P^{-1}=-N_{\rho},\quad P (-1)^{N_n} P^{-1}=(-1)^{N_n}.
\end{aligned}
\eeq 
When $N_\rho=0$, $P$ becomes a symmetry of the Hamiltonian, which can have eigenvalues $\eta_P=\pm1$. 

Therefore, the eigenstates of our model can be classified into global charge sectors labeled by $(K_{\text{tot}}, N_\rho)$ when $N_\rho\neq0$, and $(K_{\text{tot}}, N_\rho,\eta_P)$ when $N_\rho=0$.

\begin{figure}[!tbp]
\centering\includegraphics[width=0.48\textwidth]{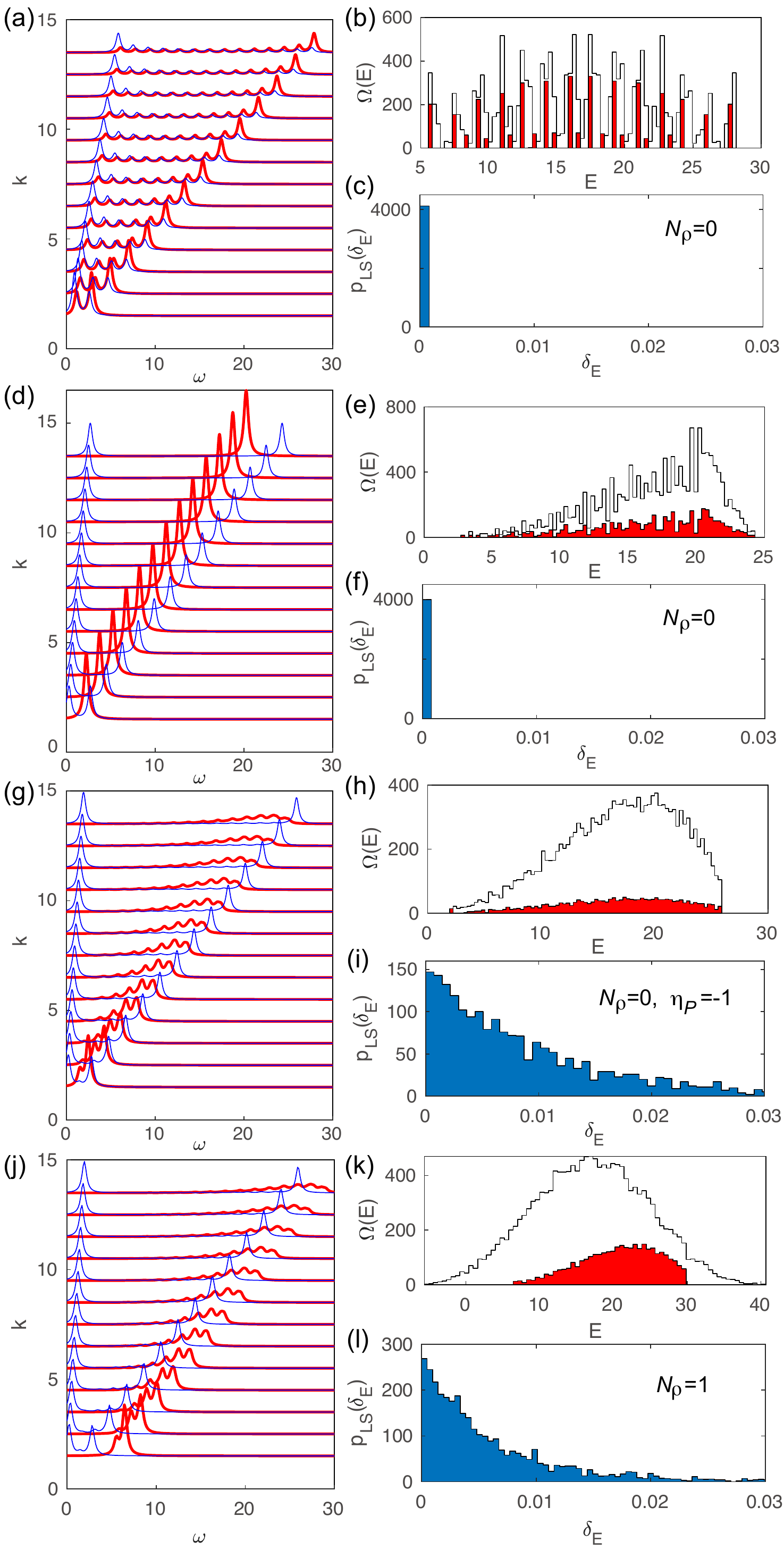}\caption{ED results up to total momentum $K_{\text{tot}}=13.5$ for different parameters. We fix $v_\rho=1.5,v_n=1$, and the other parameters are $(a)$-$(c)$: $J=0,v_{n\rho}=0.8,\mu_\rho=0$; $(d)$-$(f)$: $J=0.8\pi,v_{n\rho}=0,\mu_\rho=0$; $(g)$-$(i)$; $J=0.8\pi,v_{n\rho}=0.4,\mu_\rho=0$, and $(j)$-$(l)$: $J=0.8\pi,v_{n\rho}=0.4, \mu_\rho=4$. $(a),(d),(g), (j)$ are spectral weight of fermions $\rho$ (thick red lines) and $n$ (thin blue lines). The upper unfilled (lower filled) lines in $(b),(e),(h)$ show the total (the $N_\rho=0$ subsector) DOS in the $K_{\text{tot}}=13.5$ sector. The upper unfilled (lower filled) lines in $(k)$ show the total (the $N_\rho=1$ subsector) DOS in the $K_{\text{tot}}=13.5$ sector. $(c),(f)$ show the LSS of the $(K_{\text{tot}}=13.5,N_\rho=0)$ sector, in which cases the model is free. $(i)$ shows the LSS of the $(K_{\text{tot}}=13.5,N_\rho=0,\eta_P=-1)$ sector. $(l)$ shows the LSS of the $(K_{\text{tot}}=13.5,N_\rho=1)$ sector. }\label{fig:ED}
\end{figure}

Noting that the $\mu_\rho$ term only shifts the energy spectrum of a global charge sector by a constant $\mu_\rho N_\rho$, thus we will not discuss much on the effect of $\mu_\rho$. But $\mu_\rho$ does affect the spectral weights, which can be seen from Fig. \ref{fig:ED} and App. \ref{App:D}. 2. Fig. \ref{fig:ED} shows a comparison among the numerical results of the $J=0$ case ((a)-(c)), the $v_{n\rho}=0$ case ((d)-(f)), and the generic case with both $J$ and $v_{n\rho}$ nonzero ((g)-(i) for $\mu_\rho=0$, and (j)-(l) for $\mu_\rho=4$), with the total momentum $K_{\text{tot}}$ up to $\frac{27}{2}$, and the parameters are given in the Fig. \ref{fig:ED} caption. In particular, we compute the many-body level spacing statistics (LSS) in each global charge sector, which is known to obey a Poisson distribution $p_{LS}(s)\propto e^{-s/s_0}$ if the model is integrable \cite{Berry1977}, and obey a Wigner-Dyson distribution $p_{LS}(s)\propto s^m e^{-s^2/s_0^2}$ ($m=1,2,4$ for Gaussian orthogonal, unitary and symplectic ensembles, respectively) if the model is fully chaotic in the charge sector \cite{Bohigas1984,Dyson1970,Wigner1967}.

In the $J=0$ case, Fig. \ref{fig:ED}(a) shows the spectral weights $A_\rho(\omega,k)$ of fermion $c_\rho'$ (red thick lines) and $A_{n}(\omega,k)$ of fermion $c_n'$ (blue thin lines) from ED, which matches well with the theoretical curves in Fig. \ref{fig:sw}(a) derived from the free boson picture. With a finite system size, the numerical spectral weights are a set of discrete delta functions, which we have broadened into Lorentzian functions (App. \ref{App:D}. 2). In Fig. \ref{fig:ED}(b), we plot the density of states (DOS) at total momentum $K_{\text{tot}}=\frac{27}{2}$ of all the $N_\rho$ sectors (black unfilled line) and of only the $N_\rho=0$ sector (red filled line). Fig. \ref{fig:ED}(c) shows the LSS in the $K_{\text{tot}}=\frac{27}{2},N_\rho=0$ sector, which is approximately a delta function (or Poisson distribution with an infinite decay exponent) because of the extensive level degeneracy of free bosons.

In the $v_{n\rho}=0$ case, $c_\rho'$ and $c_n'$ are free fermions, and their ED spectral weights are delta functions as shown in Fig. \ref{fig:ED}(d), agreeing with the theoretical plot in Fig. \ref{fig:sw}(d). Similar to the $J=0$ case, the existence of enormous degenerate levels (Fig. \ref{fig:ED}(e)) leads to an approximate delta function LSS in each conserved ($K_{\text{tot}},N_\rho$) sector, as shown in Fig. \ref{fig:ED}(f).

In the generic $J\neq0$, $v_{n\rho}\neq0$ case, the spectral weights of fermions $c_n'$, $c_\rho'$ show behaviors in between the free boson and free fermion pictures, as shown in Fig. \ref{fig:ED}(g) and (j) (see more examples in App. \ref{App:D}. 2). Accordingly, the DOS (Fig. \ref{fig:ED}(h),(k)) is smoother than that in the free boson and free fermion cases (Fig. \ref{fig:ED}(b),(e)). Most intriguingly, the LSS of each conserved global charge sector shows a Poisson statistics. Fig. \ref{fig:ED}(i) shows the LSS in the $(K_{\text{tot}}=\frac{27}{2},N_\rho=0,\eta_P=-1)$ sector, and Fig. \ref{fig:ED}(l) shows the LSS of the  ($K_{\text{tot}}=\frac{27}{2},N_\rho=1$) sector. More examples are given in App. \ref{App:D}. 2 Fig. (\ref{figS:LSS}). Note that a nonzero $\mu_\rho$ in Eq. (\ref{eq:LaF2}) will not change the LSS, which only globally shifts the energy of each sector. This suggests that there are more hidden local conserved quantities (local symmetries), or the generic model in Eq. (\ref{eq:LaF2}) may in fact be quantum integrable if there are sufficient number of hidden local conserved quantities, which is an intriguing future problem. We note that the Poisson LSS in the generic case is still drastically different from the delta-function like LSS in the free boson and free fermion cases. This indicates that randomness exists in the energy spectrum of the generic case,  unlike the free cases which yield extensively degenerate many-body energy levels.

\section{The effect of irrelevant terms}

Our model has been ignoring the irrelevant terms so far. Generically, if a model is integrable at low energies, irrelevant terms may lead to breaking of integrability at high energies. Examining such integrability breaking by irrelevant terms would also support that the Poisson LSS shown in our edge model is not an artefact of numerical calculation.

In this section, we introduce an irrelevant nonlinear term to the Lagrangian density of our edge model, which preserves all the global symmetries we have identified:
\beq\label{eq-non}
\mathcal{L}_{\text{non}}=-i\lambda\left(c^\dag_{\rho}\partial_x^3c_{\rho}+c^\dag_{n}\partial_x^3c_{n}\right)\,,
\eeq 
where $\lambda$ is the coupling strength. This term gives an additional nonlinear dispersion relation $\lambda k^3$ for the free part of both the charged and neutral fermions. A simple dimension counting reveals that this nonlinear term is irrelevant and thus has vanishing effects on the low energy physics. To examine the effect of this term in Eq. (\ref{eq-non}) at high energies, one can either go to larger momentum $K_{\text{tot}}$ or larger coupling strength $\lambda$, which are equivalent up to a scaling transformation. Numerically, it is easier to enlarge the coupling $\lambda$ than the momentum $K_{\text{tot}}$. 

It has been shown that by adding such nonlinear terms to the free chiral boson Luttinger liquid theory, the integrability will be broken at high energies as revealed by the LSS \cite{Biao2022,Frank2022}. Therefore, we expect a similar breaking of the potential integrability of our model by such irrelevant nonlinear terms here.

\begin{figure}[tbp]
\centering\includegraphics[width=0.48\textwidth]{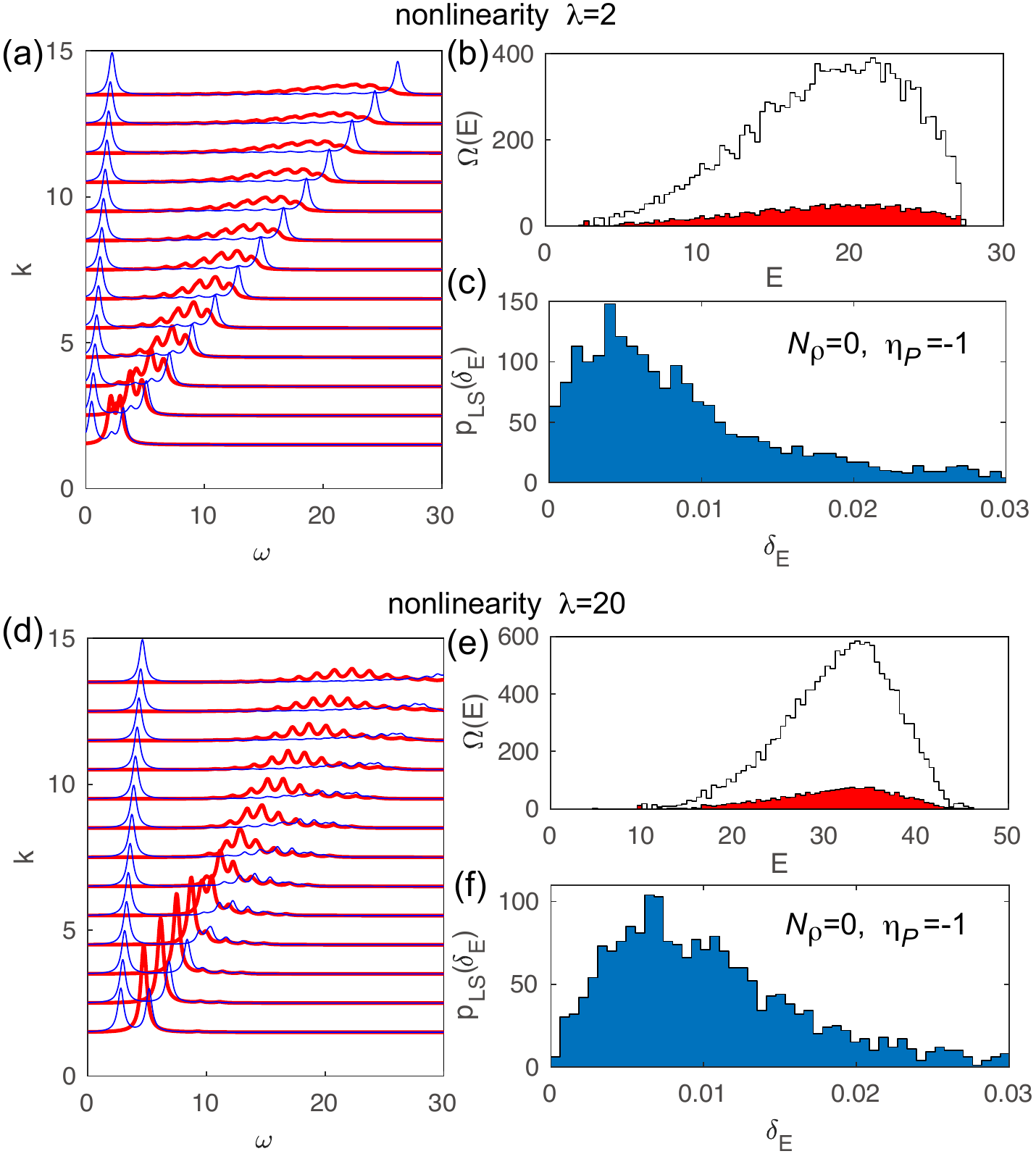}\caption{ED results up to total momentum $K_{\text{tot}}=13.5$ with an irrelevant nonlinear cubic dispersion term (with coefficient $\lambda$, see Eq. (\ref{eq-non})) added, and the rest parameters are $v_\rho=1.5,v_n=1,\mu_\rho=0, J=0.8\pi,v_{n\rho}=0.4$, and $\mu_\rho=0$ (the same as Fig. \ref{fig:ED}(g)). We set $\lambda=2$ ($\lambda=20$) in panels $(a)-(c)$ ($(d)-(f)$). Panels $(a),(d)$ show spectral weight of fermions $\rho$ (thick red lines) and $n$ (thin blue lines). The upper (lower) lines in $(b),(e)$ show the total (the $N_\rho=0,\eta_P=-1$ subsector) DOS in the $K_{\text{tot}}=13.5$ sector. $(c),(f)$ show the LSS of the $(K_{\text{tot}}=13.5,N_\rho=0,\eta_P=-1)$ sector. When the nonlinearity $\lambda$ increases, the LSS in each charge sector evolves from Poisson (Fig. \ref{fig:ED}(g)) to Wigner-Dyson (GOE) (this figure).} \label{fig:EDnon}
\end{figure}

As shown in Fig.~\ref{fig:EDnon} (as compared with Fig.~\ref{fig:ED}(i)), which are calculated from ED (see details in App. \ref{App:D}. 2), when the coupling strength $\lambda$ increases, the LSS in a global charge sector clearly shows a crossover from Poisson distribution to a Gaussian orthogonal ensemble (GOE) Wigner-Dyson distribution. The GOE type Wigner-Dyson LSS is due to the combined anti-unitary $\mathcal{PT}$ symmetry of our model, where $\mathcal{P}$ is the spatial inversion and $\mathcal{T}$ is the spinless time-reversal symmetry. This indicates that the irrelevant nonlinear term in Eq. \ref{eq-non} makes our model quantum chaotic at high energies, in contrast to the potentially integrable behavior at low energies. This also indicates that any hidden local conserved charges in each global charge sector will be broken by the irrelevant terms.

\section{Discussion} We have shown that the generic $4/3$ FQH edge model is equivalent to an interacting fermion model with a $p$-wave pairing in the neutral channel (Eq. (\ref{eq:LaF2})). It has two solvable points as free chiral bosons and free chiral (Majorana) fermions, respectively. It would be interesting to tune the system towards $v_{n\rho}=0$ and momentum difference $p=0$ (by tuning the edge potentials and chemical potential), where three different velocities of free chiral fermions (one charge $\frac{2e}{\sqrt{3}}$ and two Majorana) emerge. For generic parameters, each conserved global charge and momentum sector shows a Poisson LSS in our ED calculation, suggesting the existence of more hidden local conserved quantities. This also suggests the possibility that the generic model is quantum integrable. On the other hand, adding irrelevant nonlinear kinetic terms shifts the LSS from Poisson to Wigner-Dyson at high energies, which also indicates that any hidden local conserved quantities we have not identified yet are only exact at low energies. Moreover, it will be interesting and important to understand if the hidden local conserved quantities could protect the quantum coherence and obstruct the thermalization of the $4/3$ FQH edge states, in contrast to quantum chaotic low-energy chiral edge theories \cite{lian2019,hu2021chiral} where edge states are decoherent and fast scrambling. Further practical questions include if interference of different edges can be observed for the $4/3$ FQH chiral edge state, similar to that of the $1/3$ FQH state which was employed to measure the anyon braiding \cite{Nakamura2020}.

There are more FQH edge theories which exhibit a similar physics. For example, in the experimentally accessible regimes, the edge theory of the $(3,3,1)$ Halperin state \cite{Halperin1983,Halperin1984} at total filling $\nu=\frac{1}{2}$ also allows only one charge-conserving tunneling term $Je^{ipx} e^{i\bm{l}_{(3,3,1)}^T\bm{\phi}}$ with dimensionless coupling $J$ similar to Eq. (\ref{Lag1}), where $\bm{l}_{(3,3,1)}=(2,-2)^T$, and no other tunneling term is more relevant. The other example is the $(1,7,2)$ Halperin state at total filling $\nu=\frac{4}{3}$, which allows only one tunneling term of dimensionless coupling with $\bm{l}_{(1,7,2)}=(3,7)^T$. More series of such Abelian bilayer FQH states are identified in App. \ref{App:E}. All of these models can be mapped into the fermion model in Eq. (\ref{eq:LaF}) and thus exhibit similar physics. An interesting future direction is to study chiral edge models allowing multiple tunneling terms with dimensionless couplings, which may not be mapped to a fermion model. One may ask if the competition between different tunneling terms (which are equally relevant) will bring testable robust quantum chaos at low energies.

Lastly, in our numerical ED calculations, we considered only pure FQH edges without any disorder. Disorder is certainly a significant aspect of FQH physics, however, it is difficult to be added to numerical calculations due to its momentum conservation breaking nature. Therefore, we leave it for future studies from more analytical perspectives. For a clean enough sample, our theoretical analysis with spatially uniform coupling $J$ and chemical potential $\mu_\rho$ can be regarded as a lowest order approximation keeping the mean values of the disordered potentials, which are more relevant than their fluctuations. Moreover, in FQH edge interference experiments \cite{Nakamura2020}, to maintain coherence, the length of edge is usually not too long. Disorder effects are thus expected to play less of a role in such conditions.  

\emph{Acknowledgments.} We thank for the helpful discussions with Duncan Haldane, Shivaji Sondhi and Bertrand Halperin. This work is supported by the Alfred P. Sloan Foundation, and NSF through the Princeton University’s Materials Research Science and Engineering Center DMR-2011750. Additional support was provided by the Gordon and Betty Moore Foundation through Grant GBMF8685 towards the Princeton theory program.

\bibliographystyle{apsrev4-1}
\bibliography{main}


\appendix
\onecolumngrid

\section{Fermionization of edge theory for the $\nu=4/3$ Quantum Hall state}\label{App:A}
The Lagrangian density describing the edge theory of the Abelian $\nu=4/3$ quantum Hall state is
\beq 
\mathcal{L}=-\frac{1}{4\pi}\sum_{i,j=1,2}\partial_x \phi_i(K_{ij}\partial_t+V_{ij}\partial_x)\phi_j
\eeq with K-matrix $K=\begin{pmatrix}
3 & 0\\
0 & 1 \\
\end{pmatrix}$, charge vector $\bm{q}=(1,1)^T$ and $V_{ij}$ is the real symmetric velocity matrix. For spatially uniform tunneling between these two edge modes, the leading interaction term is
\beq
-J(x)e^{3i\phi_1-i\phi_2}-h.c., \quad J(x)=Je^{ipx}\ ,
\eeq where a momentum difference $p$ exist between the two modes adding a $x$-dependent phase factor in the coupling strength $J(x)$. The hopping operator $e^{3i\phi_1-i\phi_2}$ has scaling dimension two which renders the coupling constant $J$ dimensionless and non-negligible at low energies.

Our Lagrangian now becomes 
\beq \label{seq:cLag}
\mathcal{L}=-\frac{1}{4\pi}\sum_{i,j=1,2}\partial_x \phi_i(K_{ij}\partial_t+V_{ij}\partial_x)\phi_j-(Je^{ipx}e^{3i\phi_1-i\phi_2}+h.c.)
\eeq

We can further separate the neutral degrees of freedom from the charge degrees of freedom by the following basis transformation 
\beq
\phi_n=\frac{3\phi_1-\phi_2}{2},\quad \phi_{\rho}=\frac{\sqrt{3}(\phi_1+\phi_2)}{2}
\label{seq:nrho}
\eeq
characterized by vectors (transformation coefficients) $\bm{l}_n=(\frac{3}{2},-\frac{1}{2})$ and $\bm{l}_\rho=(\frac{\sqrt{3}}{2},\frac{\sqrt{3}}{2})$. Fermionizing as $c_{n}=e^{i\phi_{n}}$ and $c_{\rho}=e^{i\phi_{\rho}}$, we obtain an equivalent interacting fermion model:
\beq
\label{seq:LaF}
\mathcal{L}=ic^\dagger_n(\partial_t+v_n\partial_x)c_n+ic_\rho^\dagger(\partial_t+v_\rho \partial_x)c_\rho+\left(\frac{J e^{ipx}}{2\pi}ic_n \partial_x c_n+h.c.\right)-2\pi v_{n\rho} c^\dagger_n c_n c^\dagger_\rho c_\rho\ .
\eeq where electric charges $Q_n=0$ and $Q_\rho=\frac{2e}{\sqrt{3}}$ according to $Q_{\bm{l}}=\bm{l}^TK^{-1}\bm{t}$. The velocities are given by $v_\rho=\frac{V_{11}+6V_{12}+9V_{22}}{12}$, $v_n=\frac{V_{11}-2V_{12}+V_{22}}{4}$ and $v_{n\rho}=\frac{V_{11}+2V_{12}-3V_{22}}{4\sqrt{3}}$. 
We have taken the following convention for bosonization mappings with a point-splitting procedure in the $x$-direction, the details of which can be found in Appendices of Refs. \cite{lian2019,hu2021chiral}:
\beq
c^\dag_\alpha c_\alpha =\frac{\partial_x\phi_\alpha}{2\pi},\quad -ic^\dag_\alpha\partial_x c_\alpha =\frac{(\partial_x\phi_\alpha)^2}{4\pi},\quad -ic_\alpha\partial_x c_\alpha =2\pi e^{2i\phi_\alpha}, \quad (\alpha=n,\rho)\ .
\eeq

\section{Solvable point at $J=0$}\label{App:B}
When $J=0$, Eq.~(\ref{seq:cLag}) becomes a free boson model
\beq
\mathcal{L}=-\frac{1}{4\pi}\partial_x\phi_\rho(\partial_t+v_\rho\partial_x)\phi_\rho-\frac{1}{4\pi}\partial_x\phi_n(\partial_t+v_n\partial_x)\phi_n-\frac{2}{4\pi} v_{n\rho}\partial_x \phi_n \partial_x \phi_\rho
\eeq 
Let $\zeta=\arctan \frac{2v_{n\rho}}{v_\rho-v_n}$, we perform the following basis transformation for boson fields
\beq
\phi'_-=-\sin{\frac{\zeta}{2}}\phi_\rho+\cos{\frac{\zeta}{2}}\phi_n, \quad \phi'_+=\cos{\frac{\zeta}{2}}\phi_\rho+\sin{\frac{\zeta}{2}}\phi_n\ .
\eeq

This diagonalizes the above Lagrangian density and gives us
\beq
\mathcal{L}=-\frac{1}{4\pi}\sum_{\eta=\pm}\partial_x \phi'_\eta(\partial_t+v'_\eta \partial_x) \phi'_\eta
\eeq with $v'_\eta=\frac{v_n+ v_\rho\pm \sqrt{4v_{n\rho}^2+(v_n-v_\rho)^2}}{2}$ and $\la \phi'_\eta (t,x) \phi'_{\eta'}(0,0) \ra=-\delta_{\eta\eta'}\log{[2\pi i(v'_\eta t-x-i0^+)]}$ up to constant shifts of boson fields $\phi'_\eta$.
Following from this, we have
\beq
\bea
&\la\phi_\rho(t,x) \phi_\rho(0,0) \ra=\sin^2{\frac{\zeta}{2}}\la \phi'_- (t,x) \phi'_{-}(0,0) \ra+\cos^2{\frac{\zeta}{2}}\la \phi'_+ (t,x) \phi'_{+}(0,0) \ra\\
&\la\phi_n(t,x) \phi_n(0,0) \ra=\cos^2{\frac{\zeta}{2}}\la \phi'_- (t,x) \phi'_{-}(0,0) \ra+\sin^2{\frac{\zeta}{2}}\la \phi'_+ (t,x) \phi'_{+}(0,0) \ra\, 
\eea
\eeq
and thus (noting that $\sin^2{\frac{\zeta}{2}}=\frac{1-\cos\zeta}{2}$ and $\cos^2{\frac{\zeta}{2}}=\frac{1+\cos\zeta}{2}$)
\beq
\bea
&G_\rho(t,x)=\la c_\rho(t,x) c_\rho^\dag(0,0)\ra=\frac{1}{2\pi i}\frac{1}{(v'_- t-x-i0^+)^{\frac{1-\cos\zeta}{2}}(v'_+ t-x-i0^+)^{\frac{1+\cos\zeta}{2}}}\\
&G_n(t,x)=\la c_n(t,x) c_n^\dag(0,0)\ra=\frac{1}{2\pi i}  \frac{1}{(v'_- t-x-i0^+)^{\frac{1+\cos\zeta}{2}} (v'_+ t-x-i0^+)^{\frac{1-\cos\zeta}{2}}}\ .
\eea
\eeq 
From Eq.~(\ref{seq:nrho}), we also have
\beq
\bea
&\la\phi_1(t,x) \phi_1(0,0) \ra=\frac{2+\cos{\zeta}-\sqrt{3}\sin{\zeta}}{12}\la\phi'_-(t,x) \phi'_-(0,0)\ra+\frac{2-\cos{\zeta}+\sqrt{3}\sin{\zeta}}{12}\la\phi'_+(t,x) \phi'_+(0,0)\ra\\
&\la\phi_2(t,x) \phi_2(0,0) \ra=\frac{2-\cos{\zeta}+\sqrt{3}\sin{\zeta}}{4}\la\phi'_-(t,x) \phi'_-(0,0)\ra+\frac{2+\cos{\zeta}-\sqrt{3}\sin{\zeta}}{4}\la\phi'_+(t,x) \phi'_+(0,0)\ra\ ,
\eea
\eeq
and thus the charge $\frac{e}{3}$ anyon $\chi=e^{i\phi_1}$ and the charge $e$ electron $\psi=e^{i\phi_2}$ have two-point functions
\beq
\bea
&G_\chi(t,x)=\la \chi(t,x)\chi^\dag(0,0)\ra 
=\frac{1}{(2\pi i)^{\frac{1}{3}}}\frac{1}{(v'_- t-x-i0^+)^{\frac{2+\cos{\zeta}-\sqrt{3}\sin{\zeta}}{12}}(v'_+ t-x-i0^+)^{\frac{2-\cos{\zeta}+\sqrt{3}\sin{\zeta}}{12}}}\\
&G_\psi(t,x)=\la \psi(t,x)\psi^\dag(0,0)\ra 
=\frac{1}{2\pi i}\frac{1}{(v'_- t-x-i0^+)^{\frac{2-\cos{\zeta}+\sqrt{3}\sin{\zeta}}{4}}(v'_+ t-x-i0^+)^{\frac{2+\cos{\zeta}-\sqrt{3}\sin{\zeta}}{4}}}\ .
\eea
\eeq

For charge and neutral fermons $c_\rho$ and $c_n$, and the original electron operator $\psi=e^{i\phi_2}$, we can work out their retarded Green's functions and spectral weight functions. To this end, we work in Euclidean spacetime with imaginary time $(\tau=it)$ for calculation conveniences. Their Green's function in momentum space is
\beq
\bea
&G_\alpha(i\omega_\tau,k)=\int_{-\infty}^{\infty} d\tau e^{-i\omega_\tau \tau} \int^{\infty}_{-\infty} dx e^{-ikx}\frac{1}{2\pi}(v'_-\tau-ix)^{-1+\sigma_\alpha} (v'_+\tau-ix)^{-\sigma_\alpha}\\
&=\frac{-\sin{\pi \sigma_\alpha}}{\pi}\int_{-\infty}^{\infty} d\tau e^{-i\omega_\tau \tau}\int^{v'_1\tau}_{v'_2\tau} dy e^{-ky}(y-v'_-\tau)^{-1+\sigma_\alpha} (v'_+\tau-y)^{-\sigma_\alpha} \quad (y=ix)
\eea
\eeq where $\alpha=n,\rho,\psi$. This integration can be further simplified to
\beq
\bea
&=\frac{-\sin{(\pi \sigma_\alpha)}}{\pi }\int_0^{\infty} dy_+ y_+^{-\sigma_\alpha} e^{-k_- y_+}\int^{\infty}_{0} dy_- y_-^{-1+\sigma_\alpha}e^{-k_+y_-} 
&=\frac{1}{(i\omega_\tau+v'_- k)^{1-\sigma_\alpha}(i\omega_\tau+v'_+ k)^{\sigma_\alpha}}
\eea
\eeq by changing the variables of integration $y_-=\frac{y-v'_-\tau}{v'_+-v'_-}$ and $y_+=\frac{v'_+\tau-y}{v'_+-v'_-}$. We have $k_\pm=v'_{\pm}k+i\omega_\tau$ and assume $\text{Re}~k_\pm>0$.
Analytically continuing $i\omega_\tau\rightarrow -\omega-i0^+$, we find the retarded Green's function:
\beq
G_{\alpha,\text{ret}}(\omega,k)=\frac{1}{(v'_- k-\omega-i0^+)^{1-\sigma_\alpha}(v'_+ k-\omega-i0^+)^{\sigma_\alpha}}
\eeq The spectral weight is then
\beq
\label{seq:Anr}
A(\omega, k)=2\text{Im} G_{\alpha,\text{ret}}(\omega,k)=\frac{2\sin{(\pi\sigma_\alpha)}\Theta(v'_+ k-\omega)\Theta(\omega-v'_- k)}{(v'_+k-\omega)^{\sigma_\alpha}(\omega-v'_-k)^{1-\sigma_\alpha}}
\eeq such that
\beq
1=\int_{-\infty}^{\infty}\frac{d\omega}{2\pi} A(\omega,k) \ .
\eeq 
For our cases, $\sigma_\rho=\frac{1+\cos\zeta}{2}$, $\sigma_n=\frac{1-\cos\zeta}{2}$, and $\sigma_\psi=\frac{2+\cos{\zeta}-\sqrt{3}\sin{\zeta}}{4}$.

\section{Solvable point at $v_{n\rho}=0$}\label{App:C}
\subsection{Two-point functions for fermions with generic momentum difference $p$}
When $v_{n\rho}=0$, the neutral $n$ and charge $\rho$ fermions in Eq.~(\ref{seq:LaF}) are completely decoupled. Let us define a new fermion operator $c'_n=e^{\frac{ipx}{2}}c_n$, such that our Hamiltonian density becomes position $x$ independent:
\beq
H=\int dx [-iv_\rho c_\rho^\dagger  \partial_x c_\rho-iv_n c'^\dagger_n\partial_x c'_n-\frac{J}{2\pi}\left(ic'_n \partial_x c'_n+h.c.\right)+\frac{v_n p}{2}c'^\dagger_n c'_n]\ .
\eeq 

Fourier transforming to momentum space, we have
\beq
H=\int_{-\infty}^{\infty} dk [v_\rho k c^\dagger_\rho(k)c_\rho(k)+v_n (k+\frac{p}{2}) c'^\dagger_n(k)c'_n(k)-\frac{Jk}{2\pi} c'_n(k)c'_n(-k)+\frac{Jk}{2\pi}  c'^\dag_n(k)c'^\dag_n(-k)]\ .
\eeq
We see that $c_\rho$ is with linear dispersion relation $\omega_\rho=v_\rho k$. For the neutral fermion $c'_n$, we can perform a Bogoliubov transformation
\beq
\begin{pmatrix}b_{n}(k)\\b^\dag_{n}(-k)\end{pmatrix}=\begin{pmatrix} \lambda_{k_-} & \lambda_{k_+}\\ -\lambda_{k_+} & \lambda_{k_-} \end{pmatrix} \begin{pmatrix}c'_{n}(k)\\c'^\dag_{n}(-k)\end{pmatrix}
\eeq with
\beq
\lambda_{k_\pm}=\frac{2Jk}{\sqrt{(2Jk)^2+(\pi v_n p\pm\sqrt{(2Jk)^2+(\pi v_n p)^2})^2}}
\eeq which gives us the spectrum of the Bogoliubov fermion $b_{n}(k)$: 
\begin{equation}
\omega_{b_n}(k)=v_nk+\frac{1}{2\pi}\sqrt{(2Jk)^2+(\pi v_n p)^2}\ .
\end{equation}
For two-point function calculation conveniences, let us work in Euclidean spacetime with imaginary time $\tau=it$.
Since
\beq
c'_n(k)=\lambda_{k_-}b_{n}(k)-\lambda_{k_+}b^\dag_{n}(-k)\ ,
\eeq 
the imaginary time ordered two-point function is defined as
\begin{equation}
\begin{split}
&G_{n}'(-i\tau,k)=\langle T c_n'(\tau,k)c_n'^\dag (0,k) \rangle = |\lambda_{k_-}|^2\langle T b_n(\tau,k)b_n^\dag (0,k) \rangle + |\lambda_{k_+}|^2\langle T b_n^\dag(\tau,-k)b_n (0,-k) \rangle \\
&= |\lambda_{k_-}|^2\langle T b_n(\tau,k)b_n^\dag (0,k) \rangle - |\lambda_{k_+}|^2\langle T b_n (0,-k)b_n^\dag(\tau,-k) \rangle\\
&= |\lambda_{k_-}|^2 G_b(-i\tau,k) -|\lambda_{k_+}|^2 G_b(i\tau,-k)\ ,
\end{split}
\end{equation} where $T$ stands for time ordering.
Fourier transforming the time direction gives us
\begin{equation}
\begin{split}
&G_{n}'(i\omega_\tau,k)=|\lambda_{k_-}|^2 G_b(i\omega_\tau,k) -|\lambda_{k_+}|^2 G_b(-i\omega_\tau,-k)\\
&=\frac{|\lambda_{k_-}|^2}{i\omega_\tau-v_nk-\frac{1}{2\pi}\sqrt{(2Jk)^2+(\pi v_n p)^2}} +\frac{|\lambda_{k_+}|^2}{i\omega_\tau-v_nk+\frac{1}{2\pi}\sqrt{(2Jk)^2+(\pi v_n p)^2}}\\
&=\frac{i\omega_\tau-v_nk+\frac{v_n p}{2}}{(i\omega_\tau-v_nk)^2-\frac{1}{4\pi^2}[(2Jk)^2+(\pi v_n p)^2]}=\frac{i\omega_\tau-v_nk+\frac{v_n p}{2}}{(i\omega_\tau-v_{n_+}k)(i\omega_\tau-v_{n_-}k)-(\frac{v_n p}{2})^2}\ .
\end{split}
\end{equation}
where we have used 
\begin{equation}
G_b(i\omega_\tau,k)=\frac{1}{i\omega_\tau-\omega_{b_n}(k)},\quad |\lambda_{k_\mp}|^2=\frac{1}{2}\left(1\pm\frac{\pi v_n p}{\sqrt{(2Jk)^2+(\pi v_n p)^2}} \right)\ .
\end{equation}
The real space Green's function is thus
\begin{equation}
\begin{split}
&G_n'(-i\tau,x)=\int \frac{d\omega_\tau dk}{(2\pi)^2}e^{i\omega_\tau\tau+ikx}G_{n}'(i\omega_\tau,k)=\int \frac{d\omega_\tau dk}{(2\pi)^2}e^{i\omega_\tau\tau+ikx}\frac{i\omega_\tau-v_nk+\frac{v_n p}{2}}{(i\omega_\tau-v_{n_+}k)(i\omega_\tau-v_{n_-}k)-(\frac{v_n p}{2})^2}\ . \\
\end{split}
\end{equation} We can get the real-time Green's function $G'_n(t,x)$ by analytically continuing $-i\tau \rightarrow t$ and $i\omega_\tau \rightarrow -\omega-i0^+$. The Green's function of $c_n=e^{-\frac{ipx}{2}}c_n'$ in the original basis is given by
\begin{equation}
G_n(t,x)=G'_n(t,x)e^{-\frac{ipx}{2}}\ .
\end{equation}

\subsection{Two-point correlation functions for fermions and anyons at $p=0$}
In this subsection, we calculate explicitly the correlation functions of fermions and anyons when $p=0$.
\subsubsection{For fermions}
At $p=0$, the calculation of two-point functions for fermions are straightforward. Let us again start with Euclidean spacetime for calculation conveniences. For fermions in charge sector, in momentum space, their two-point function is given as
\beq
G_\rho(i\omega_\tau,k)=\frac{1}{i\omega_\tau +kv_\rho}, 
\eeq
Analytically continuing $i\omega_\tau \rightarrow -\omega-i0^+$, we get the retarded Green's function
\beq
G_{\rho,\text{ret}}(\omega,k)=\frac{1}{kv_\rho-\omega-i0^+}
\eeq which gives us the spectral weight function
\beq
\label{seq:Ar}
A_\rho(\omega,k)=2 \text{Im} G_{\rho,\text{ret}}(\omega,k)=2\pi\delta(\omega- kv_\rho)\ .
\eeq
Similarly, for fermions in neutral sector, we have the retarded Green function
\beq
G_{n,\text{ret}}(k,\omega)=\frac{1}{2}(\frac{1}{kv_{n_+}-\omega-i0^+ }+\frac{1}{kv_{n_-}-\omega-i0^+ })
\eeq 
where the velocities are given by $v_{n_\pm}=v_n\pm\frac{J}{2\pi}$. Then
\beq \label{seq:An}
A_n(k, \omega)=\pi(\delta(\omega-k v_{n_+})+\delta(\omega-k v_{n_-}))
\eeq signaling the appearance of two free Majorana fermions at $p=0$. Fourier transforming back to real time domain, we have
\beq
G_n(t,x)=\frac{1}{4\pi i}(\frac{1}{v_{n_+}t-x-i0^+}+\frac{1}{v_{n_-}t-x-i0^+}),\quad G_\rho(t,x)=\frac{1}{2\pi i}\frac{1}{v_{\rho}t-x-i0^+}\ .
\eeq 


\subsubsection{For anyons}
At $p=0$, two-point correlation functions for the charge $\frac{2e}{3}$ anyon $\chi'=e^{2i\phi_1}=e^{i(\phi_n+\frac{\phi_\rho}{\sqrt{3}})}$ can be readily calculated
\beq
\la \chi'(t,x)\chi'^\dag(0,0)\ra=G_n(t,x)G_\rho(t,x)^{1/3}=\frac{1}{2(2\pi i)^{4/3}}(\frac{1}{v_{n_+}t-x-i0^+}+\frac{1}{v_{n_-}t-x-i0^+})\frac{1}{(v_{\rho}t-x-i0^+)^{\frac{1}{3}}}
\eeq



For calculations of two-point correlation functions involving the charge $\frac{e}{3}$ anyon $\chi=e^{i\phi_1}$ and the charge $e$ electron $\psi=e^{i\phi_2}$, we adapt a similar strategy implemented in \cite{hu2021chiral}. We first enlarge our Hilbert space by making a duplicated copy of our $\nu=4/3$ Lagrangian density, after which the total Lagrangian reads
\beq
\begin{aligned}
&\mathcal{L}_{2c}=\mathcal{L}^{(+)}+\mathcal{L}^{(-)}\\
&=-\frac{1}{4\pi}\sum_{\eta=\pm}[\partial_x\phi_\rho^{(\eta)}(\partial_t+v_\rho\partial_x)\phi_\rho^{(\eta)}+\partial_x\phi_n^{(\eta)}(\partial_t+v_n\partial_x)\phi_n^{(\eta)}+4\pi J(e^{2i\phi_n^{(\eta)}}+h.c.)]\ ,
\end{aligned}
\eeq 
where for each copy $\eta=\pm$, we are in the neutral-charge basis defined as
\beq
\phi_n^{(\eta)}=\frac{3\phi_1^{(\eta)}-\phi_2^{(\eta)}}{2},\quad \phi_{\rho}^{(\eta)}=\frac{\sqrt{3}(\phi_1^{(\eta)}+\phi_2^{(\eta)})}{2}
\eeq
Next, let us define new fermion variables $f_i=e^{i\widetilde{\phi}_i}$ such that
\beq
  \phi_n^{(\eta)}=\frac{(\widetilde{\phi}_1-\widetilde{\phi}_2)+\eta(\widetilde{\phi}_3-\widetilde{\phi}_4)}{2}, \quad
  \phi_\rho^{(\eta)}=\frac{(\widetilde{\phi}_1+\widetilde{\phi}_2)+\eta(\widetilde{\phi}_3+\widetilde{\phi}_4)}{2}.
\eeq In literature \cite{shankar1981,shankar1983,Kane2020,Ryu2012,Maldacena1997,Kitaev2010,hu2021chiral}, this is known as the $SO(8)$ triality transformation. Under this transformation, our Lagrangian density becomes
\beq
\mathcal{L}_{2c}=-\frac{1}{4\pi}(\partial_t \widetilde{\bm{\phi}}^T \partial_x \widetilde{\bm{\phi}} +\partial_x \widetilde{\bm{\phi}}^T \widetilde{V} \partial_x \widetilde{\bm{\phi}}) -J(e^{i(\widetilde{\phi}_1-\widetilde{\phi}_2+\widetilde{\phi}_3-\widetilde{\phi}_4)}+e^{i(\widetilde{\phi}_1-\widetilde{\phi}_2-\widetilde{\phi}_3+\widetilde{\phi}_4)}+h.c.)
\eeq with velocity matrix
\beq
\widetilde{V}=\begin{pmatrix}
\widetilde{u}_+ & \widetilde{u}_- & 0 & 0\\
\widetilde{u}_- & \widetilde{u}_+ & 0 & 0\\
0 & 0 & \widetilde{u}_+ & \widetilde{u}_-\\
0 & 0 & \widetilde{u}_- & \widetilde{u}_+
\end{pmatrix}\ ,
\eeq $\widetilde{u}_+=\frac{v_\rho+v_n}{2}$, $\widetilde{u}_-=\frac{v_\rho-v_n}{2}$ and $\widetilde{\bm{\phi}}=(\widetilde{\phi}_1,\widetilde{\phi}_2,\widetilde{\phi}_3,\widetilde{\phi}_4)^T$. In the $f_i$ fermion basis, the Lagrangian density is
\beq
\mathcal{L}_{2c}=\sum_{s=1}^4 i f_s^\dagger (\partial_t+\widetilde{u}_+ x)f_s-2\pi \widetilde{u}_-(f_1^\dagger f_1 f_2^\dagger f_2+f_3^\dagger f_3 f_4^\dagger f_4)-J(f_1^\dagger f_2-f^\dagger_2 f_1)(f_3^\dagger f_4-f^\dagger_4 f_3).
\eeq This interacting fermion model can be further bosonized in a rotated fermion basis: $d_{1,\pm}=\frac{f_1\pm i f_2}{\sqrt{2}}=e^{i\theta_{1,\pm}}$, $d_{2,\pm}=\frac{f_3\pm i f_4}{\sqrt{2}}=e^{i\theta_{2,\pm}}$, such that all terms in the Lagrangian are bilinear in boson fields $\{\theta_{i,\pm}\}_{i=1,2}$. 
\beq
\begin{aligned}
&\mathcal{L}_{2c}=\sum_{s=1}^2 \sum_{\eta=\pm}id_{s,\eta}^\dag (\partial_t+\widetilde{u}_+ \partial_x)d_{s,\eta}-2\pi \widetilde{u}_-(d_{1,+}^\dag d_{1,+} d_{1,-}^\dag d_{1,-}+d_{2,+}^\dag d_{2,+} d_{2,-}^\dag d_{2,-})\\
&-J(d_{1,+}^\dag d_{1,+}-d_{1,-}^\dag d_{1,-})(d_{2,+}^\dag d_{2,+}-d_{2,-}^\dag d_{2,-})\\
&=-\frac{1}{4\pi}(\partial_t \bm{\theta}^T\partial_x \bm{\theta}+ \partial_x \bm{\theta}^T U\partial_x \bm{\theta})
\end{aligned}
\eeq with 
\beq
U=\begin{pmatrix}
\widetilde{u}_+ & \widetilde{u}_- & \frac{J}{2\pi} & -\frac{J}{2\pi} \\
\widetilde{u}_- & \widetilde{u}_+ & -\frac{J}{2\pi}  & \frac{J}{2\pi} \\
\frac{J}{2\pi}  & -\frac{J}{2\pi}  & \widetilde{u}_+ & \widetilde{u}_-\\
-\frac{J}{2\pi}  & \frac{J}{2\pi}  & \widetilde{u}_- & \widetilde{u}_+
\end{pmatrix}\ ,
\eeq and $\bm{\theta}=(\theta_{1,+},\theta_{1,-},\theta_{2,+},\theta_{2,-})^T$. Finally, we can diagonalize our Lagrangian density by changing to a new basis for boson fields: $\widetilde{\theta}_{\eta \eta'}=(\theta_{1,+}+\eta\theta_{1,-}+\eta'\theta_{2,+}+\eta\eta'\theta_{2,-})$,
\beq
    \mathcal{L}_{2c}=-\frac{1}{4\pi}\sum_{\eta,\eta'=\pm} \partial_x \widetilde{\theta}_{\eta \eta'} (\partial_t+u_{\eta \eta'}\partial_x)\widetilde{\theta}_{\eta \eta'}
\eeq with velocities 
\begin{equation}
u_{+\pm}=\widetilde{u}_++\widetilde{u}_-=v_\rho,\qquad  u_{-\pm}=\widetilde{u}_+-\widetilde{u}_-\pm\frac{J}{\pi}=v_n\pm\frac{J}{\pi}=v_{n_\pm}\ ,
\end{equation}
With the above exact mappings established, we have
\beq
\begin{aligned}
&\la e^{i\widetilde{\phi}_1(t,x)} e^{-i\widetilde{\phi}_1(0,0)}  \ra=\la f_1 f_1^\dagger \ra=\la \frac{1}{2}(d_{1+} d_{1+}^\dagger+d_{1-} d_{1-}^\dagger)\ra=\la \frac{1}{2} (e^{i\theta_{1,+}}e^{-i\theta_{1,+}}+e^{i\theta_{1,-}}e^{-i\theta_{1,-}})\ra\\
&=\la e^{i(\widetilde{\theta}_1+\widetilde{\theta}_2+\widetilde{\theta}_3+\widetilde{\theta}_4)/2}e^{-i(\widetilde{\theta}_1+\widetilde{\theta}_2+\widetilde{\theta}_3+\widetilde{\theta}_4)/2} \ra=\frac{1}{2\pi i}\frac{1}{(v_\rho t-x-i0^+)^{1/2}(v_{n,+} t-x-i0^+)^{1/4}(v_{n,-} t-x-i0^+)^{1/4}}
\end{aligned}
\eeq
Since the charge degrees of freedom is always a free theory, we have
\beq
\la e^{i\phi_\rho^{(i)}(t,x)/2} e^{-i\phi_\rho^{(i)}(0,0)/2}  \ra=\frac{1}{2\pi i}\frac{1}{ (v_\rho t-x-i0^+)^{1/4}}
\eeq
From the fact that $\widetilde{\phi}_1=\frac{\phi_n^{(+)}+\phi_n^{(-)}+\phi_\rho^{(+)}+\phi_\rho^{(-)}}{2}$, we have $\la e^{-i\widetilde{\phi}_1} e^{i\widetilde{\phi}_1}  \ra=\left(\la e^{i\phi_n^{(\eta)}/2} e^{-i\phi_n^{(\eta)}/2}  \ra \la e^{i\phi_\rho^{(\eta)}/2} e^{-i\phi_\rho^{(\eta)}/2}  \ra\right)^2$,  which leads to
\beq
\la e^{i\phi_n^{(\eta)}(t,x)/2} e^{-i\phi_n^{(\eta)}(0,0)/2}\ra=\frac{\sqrt{\la e^{i\widetilde{\phi}_1} e^{-i\widetilde{\phi}_1}  \ra}}{\la e^{i\phi_\rho^{(\eta)}/2} e^{-i\phi_\rho^{(\eta)}/2}  \ra}=\frac{1}{(2\pi i)^{1/4}}\frac{1}{(v_{n+} t-x-i0^+)^{1/8}(v_{n,-} t-x-i0^+)^{1/8}}\ .
\eeq
Since $\phi_1=\frac{1}{2}\phi_n^{(\eta)}+\frac{1}{2\sqrt{3}}\phi_\rho^{(\eta)}$ and  $\phi_2=\frac{\sqrt{3}}{2}\phi_\rho^{(\eta)}-\frac{1}{2}\phi_n^{(\eta)}$, this gives us
\beq
\begin{aligned}
&G_\chi(t,x)=\la \chi(t,x) \chi^\dag(0,0)  \ra=\la e^{i(\frac{1}{2}\phi_n^{(\eta)}+\frac{1}{2\sqrt{3}}\phi_\rho^{(\eta)})} e^{-i(\frac{1}{2}\phi_n^{(\eta)}+\frac{1}{2\sqrt{3}}\phi_\rho^{(\eta)})} \ra=\la e^{i\frac{1}{2}\phi_n^{(\eta)}} e^{-i\frac{1}{2}\phi_n^{(\eta)}}\ra \la e^{i\frac{1}{2\sqrt{3}}\phi_\rho^{(\eta)}} e^{-i\frac{1}{2\sqrt{3}}\phi_\rho^{(\eta)}} \ra\\
&=\frac{1}{(2\pi i)^{1/3}}\frac{1}{(v_\rho t-x-i0^+)^{1/12}(v_{n_+} t-x-i0^+)^{1/8}(v_{n_-}t-x-i0^+)^{1/8}}\ ,\\
&G_\psi(t,x)=\la \psi(t,x) \psi^\dag(0,0)  \ra=\la e^{i(-\frac{1}{2}\phi_n^{(\eta)}+\frac{\sqrt{3}}{2}\phi_\rho^{(\eta)})} e^{-i(-\frac{1}{2}\phi_n^{(\eta)}+\frac{\sqrt{3}}{2}\phi_\rho^{(\eta)})} \ra=\la e^{-i\frac{1}{2}\phi_n^{(\eta)}} e^{i\frac{1}{2}\phi_n^{(\eta)}}\ra \la e^{i\frac{\sqrt{3}}{2}\phi_\rho^{(\eta)}} e^{-i\frac{\sqrt{3}}{2}\phi_\rho^{(\eta)}} \ra\\
&=\frac{1}{2\pi i}\frac{1}{(v_\rho t-x-i0^+)^{3/4} (v_{n_+} t-x-i0^+)^{1/8}(v_{n_-} t-x-i0^+)^{1/8}}\ .
\end{aligned}
\eeq 
\section{Generic cases }\label{App:D}
In this section, we consider generic cases of edge theory (Eq.~(\ref{seq:cLag})) for the $\nu=4/3$ quantum Hall state with $J\neq 0$ and $v_{n\rho}\neq 0$.
\subsection{The extra phase from momentum difference $p$}
The extra phase factor $e^{ipx}$ in Eq.~(\ref{seq:cLag}) can be eliminated by performing an unitary transformation on boson fields such that we redefine them as 
\beq
\phi_n'=\phi_n+\frac{p}{2}x, \quad \phi'_\rho=\phi_\rho-\frac{v_n p}{2v_{n\rho}}x \quad\text{ and }\quad c'_n=e^{i\frac{p}{2}x}c_n,\quad  c'_\rho=e^{-i\frac{v_n p}{2v_{n\rho}}x} c_\rho\ .
\eeq
The price we pay is an additional chemical potential term $\partial_x \phi'_\rho$ with coefficient $\mu_\rho=\left(\frac{v_nv_\rho}{v_{n\rho}}-v_{n\rho}\right)p$ for the charge fermion $c'_\rho$ generated from the above transformation. Putting these together, our Lagrangian density becomes
\beq
\mathcal{L}=ic_n'^\dagger(\partial_t+v_n\partial_x)c_n'+ic_\rho'^\dagger(\partial_t+v_\rho \partial_x)c_\rho'-\mu_\rho c_\rho'^\dagger c_\rho'+\left(i\frac{J}{2\pi}c_n' \partial_x c_n'+h.c.\right)-2\pi v_{n\rho} c_n'^\dagger c_n' c_\rho'^\dagger c_\rho'\ .
\eeq 
Green's functions of both the neutral $n$ and charged $\rho$ fermions acquire pure phase factors in the new basis as a consequence:
\beq
\label{seq:Green}
G'_n(t,x)=G_n(t,x)e^{i\frac{px}{2}},\quad G'_\rho(t,x)=G_\rho(t,x)e^{-i\frac{v_npx}{2v_{n\rho}}}\ .
\eeq This change does not alter results regarding level spacing statistics (LSS), and only shift the spectral weight functions of $n$ and $\rho$ fermions by a constant momentum.

\subsection{Discretization and exact diagonlization}
In this subsection, we discuss discretization procedures and other technical details for our exact diagonlization (ED) calculations for generic cases of our $4/3$ FQH chiral edge theory. 
\begin{figure}[htbp]
\centering\includegraphics[width=0.63\textwidth]{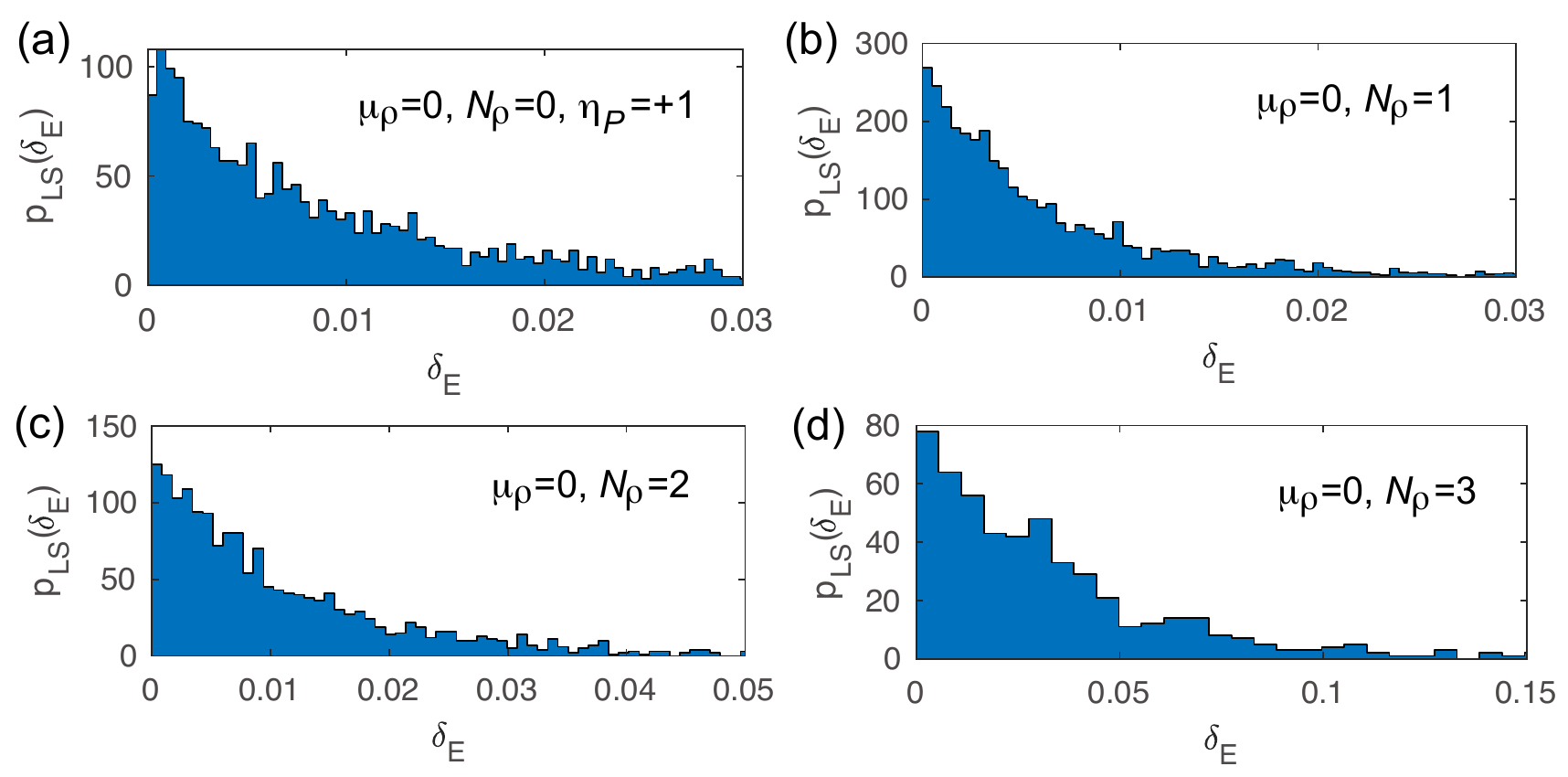}
\caption{Additional ED results for level spacing statistics (LSS) in various charge sectors at total momentum $K_{\text{tot}}=13.5$ with $J=0.8\pi,v_{n\rho}=0.4,v_\rho=1.5,v_n=1$, while the chemical potential $\mu_\rho=0$. There is an additional particle-hole charge $\eta_P=\pm1$ in the subspace of charge $N_\rho=0$. (a)-(d) are plots for charge sectors with $N_\rho=0,\eta_P=-1$ and $N_\rho=1, 2, 3$, respectively.} \label{figS:LSS}
\end{figure}

\begin{figure}[htbp]
\centering\includegraphics[width=0.9\textwidth]{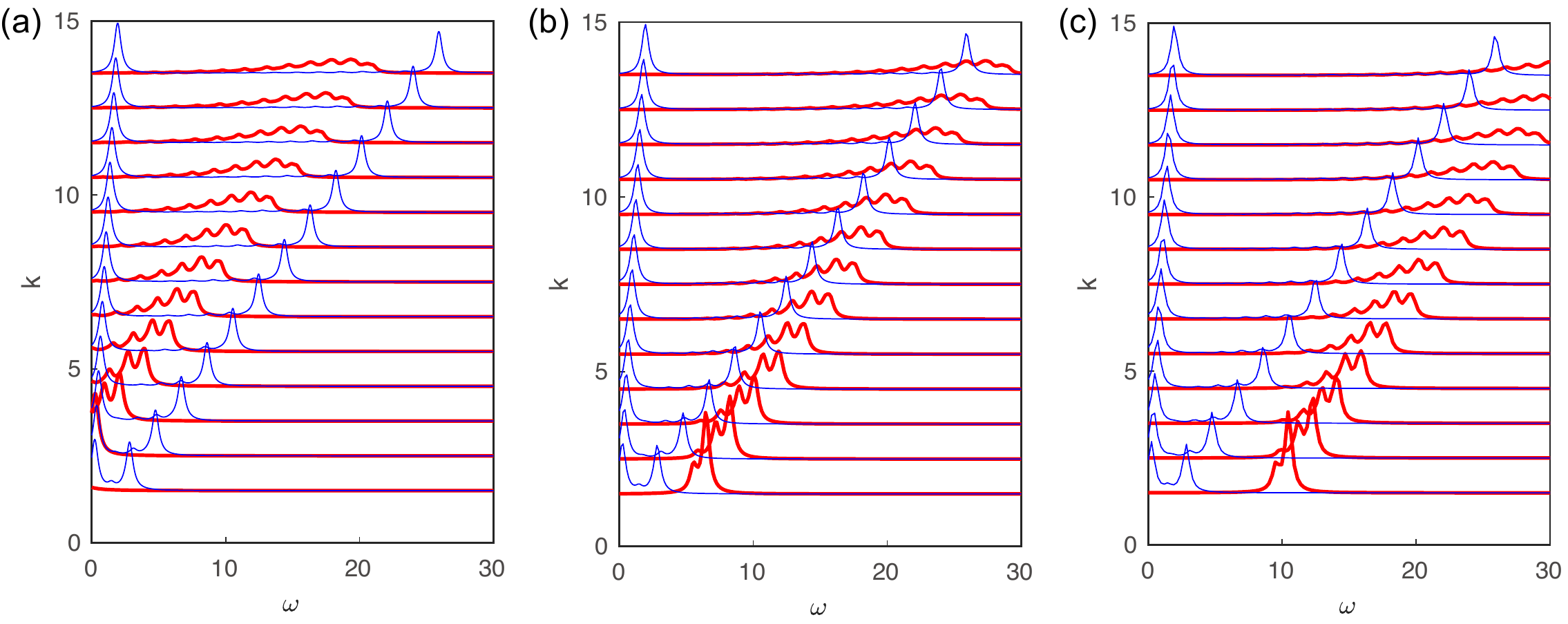}
\caption{Spectral weights of $n$ fermion (blue thin line) and $\rho$ fermion (red thick line) up to total momentum $K_{\text{tot}}=13.5$. The parameters are $J=0.8\pi,v_{n\rho}=0.4,v_\rho=1.5,v_n=1$ for all cases but with different chemical potential: (a) $\mu_\rho=-4$, (b) $\mu_\rho=4$, and (c) $\mu_\rho=8$.}\label{figS:SWmu}
\end{figure}

\begin{figure}[htbp]
\centering\includegraphics[width=\textwidth]{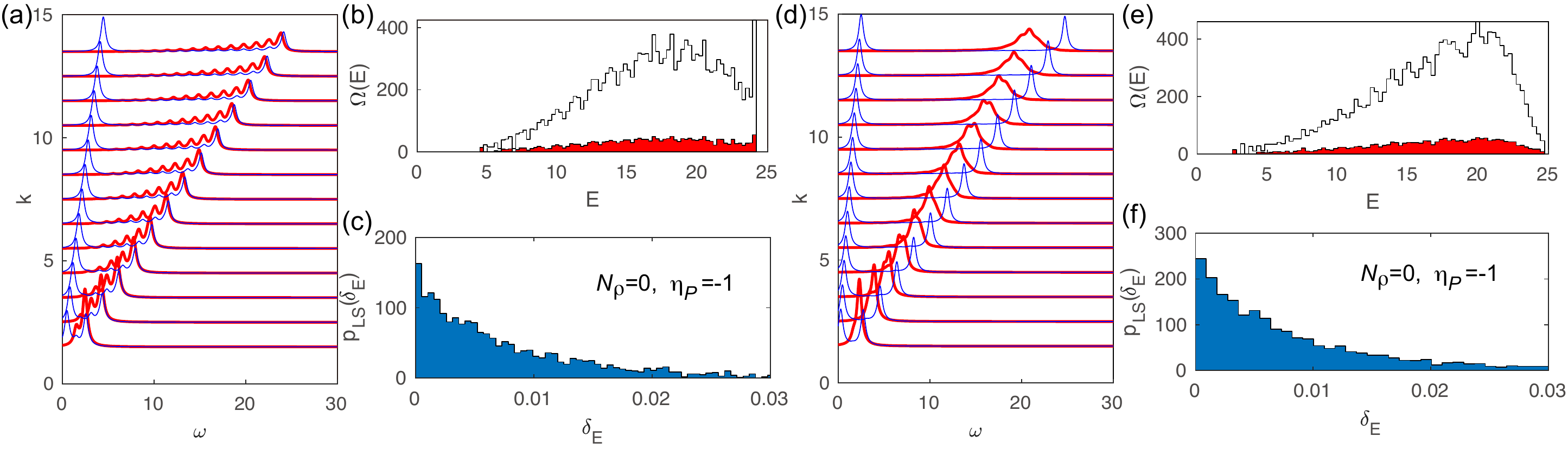}
\caption{More ED results up to total momentum $K_{\text{tot}}=13.5$. We fix $v_\rho=1.5,v_n=1,\mu_\rho=0$, while the other parameters are given by (a)-(c): $J=0.6\pi,v_{n\rho}=0.4$; (d)-(f): $J=0.8\pi,v_{n\rho}=0.2$.
$(a),(d)$ are the spectral weight of fermions $\rho$ (thick red lines) and $n$ (thin blue lines). The upper (lower) lines in $(b),(e)$ show the total (the $N_\rho=0, \eta_P=-1$ subsector) DOS in the $K_{\text{tot}}=13.5$ sector. $(c),(f)$ show the LSS of the $(K_{\text{tot}}=13.5,N_\rho=0, \eta_P=-1)$ sector.}

\label{figS:moreED}
\end{figure}

We model our edge theory as a closed one dimensional chiral system with total length $L$ and anti-periodic boundary condition for fermion operators $c_n'$ and $c_\rho'$. Imposing anti-periodic boundary condition is natural when there is no flux inside the bulk. Under this set-up, we can Fourier transform the fermion operators to momentum space:
\beq
c'_{\alpha}(x)=\frac{1}{\sqrt{L}}\sum_{k} e^{ikx}{c}_{\alpha}'(k), \qquad k\in\frac{2\pi (\mathbb{Z}+\frac{1}{2})}{L}\ ,\qquad (\alpha=n,\rho)
\eeq Our Hamiltonian in the real space is
\beq
H=\int_0^L dx [-i v_n {c'}_n^\dagger \partial_x c'_n-i v_\rho {c'}_\rho^\dagger \partial_x c'_\rho+\mu_\rho {c'}_\rho^\dag {c'}_\rho-\frac{J}{2\pi}(ic'_n \partial_x c'_n+i{c'}_n^\dagger \partial_x {c'}_n^\dagger)+2\pi v_{n\rho} {c'}_n^\dag {c'}_n {c'}_\rho^\dag c'_\rho]\ .
\eeq
Without loss of generality, we set $L=2\pi$, such that the fermion momentum $k \in \mathbb{Z}+\frac{1}{2}$. The Hamiltonian of our edge theory in real space can be transformed to momentum space as
\beq
\begin{aligned}
&H=\sum_{k=-\infty}^{\infty} \left[v_n k \left({c'}_{n}^\dagger(k) c'_{n}(k)-\frac{1}{2}\right)+ (v_\rho k+\mu_\rho)\left( {c'}_{\rho}^\dagger(k) c'_{\rho}(k)-\frac{1}{2}\right)+\frac{J}{2\pi}(k c'_{n}(-k) c'_{n}(k)-k {c'}_{n}^\dagger(-k) {c'}_{n}^\dagger(k))\right]\\
&+ v_{n\rho} \sum_{k_1,k_2,k_3,k_4=-\infty}^{\infty} \delta_{k_{1}-k_{2}+k_{3}-k_{4},0} \left({c'}^\dagger_{n}(k_{1}) c'_{n}(k_{2})-\frac{1}{2}\delta_{k_{1},k_{2}}\right)\left( {c'}^\dagger_{\rho}(k_{3}) c'_{\rho}(k_{4})-\frac{1}{2}\delta_{k_{3},k_{4}}\right)\ .
\end{aligned}
\label{seq:ed}
\eeq Because of the following relations between an electron operator and a hole operator:
\beq
c'^\dag_{\alpha, e}(k)=c'_{\alpha, h}(-k),\quad c'_{\alpha, e}(k)=c'^\dag_{\alpha, h}(-k)\ ,\qquad (\alpha=n,\rho)
\label{eq:eh}
\eeq 
by rewriting fermion operators with negative momentum as their hole operators, we can convert all the momentum of fermion operators in the summand of Eq.~(\ref{seq:ed}) as positive. Thus, for a given total many-body momentum $K_\text{tot}>0$, the Hilbert space dimension is finite, allowing us to employ the ED method to obtain the eigenstates and spectrum. To this end, the ground state of our Hamiltonian is the state with zero electrons and holes for both neutral ($n$) and charge ($\rho$) sectors.
To further reduce redundant calculations, we need to identify symmetries of our Hamiltonian. We see three different conserved quantities: the total momentum $K_{\text{tot}}$, the total charge of charge fermion $N_\rho$ and the fermion parity of neutral fermions $(-1)^{N_n}$. Since $N_n \equiv (2K_{\text{tot}}+N_\rho)  \mod 2$, only two independent conserved quantities left and we can label states in our Hilbert space by which $(K_{\text{tot}},N_\rho)$ sector they belong. In the $N_\rho=0$ sectors, our model has an additional particle-hole symmetry $P$ as defined in Eq. (\ref{eq-P-def}), yielding another conserved charge $\eta_P=\pm1$ (eigenvalues of $P$). Our ED calculations for the $N_\rho=0$ sector can thus be done for a global charge sector $(K_{\text{tot}},N_\rho=0,\eta_P)$.

After we obtain the full spectrum, we can calculate the zero-temperature spectral weight function at a given momentum $k$. It is defined as 
\beq
A_{\alpha=n,\rho}(\omega,k)=\sum_{j=1}^{N_{-k}}|\bra{-k,j}c'_\alpha(k)\ket{0}|^2\delta(\omega+E_{-k,j})+\sum_{j=1}^{N_{k}}|\bra{0}c'_\alpha(k)\ket{k,j}|^2\delta(\omega-E_{k,j})\ ,
\eeq where $\ket{k,j}$ is the $j$-th many-body eigenstate (sorted in the order of increasing energy) of total momentum $K_{\text{tot}}=k$ sector of Eq.~(\ref{seq:ed}), which has energy $E_{k,j}$, and $N_k$ is the Hilbert space dimension of the total momentum $k$ sector. It satisfies
\beq
1=\int_{-\infty}^{\infty} \frac{d\omega}{2\pi}A_{\alpha=n,\rho}(\omega,k)\ .
\eeq To plot the spectral weight, we used a Lorentizan function in places for a Dirac delta function as
\beq
\delta(\omega)\rightarrow  \frac{1}{\pi}\text{Im}\frac{1}{\omega+i\eta}\ ,
\eeq
where we take $\eta=0.3$ for all our plots.

We also plot the level spacing statistics (LSS) of conserved charge sectors in Fig.~(\ref{figS:LSS}) here and in Fig. ~(\ref{fig:ED}) of the main text. For each conserved charge $S=(K_{\text{tot}},N_\rho)$ sector (or $S=(K_{\text{tot}},N_\rho=0,\eta_P)$ sector when $N_\rho=0$), assume the energy levels are $E_{S,j}$ with $j$ sorted in the energy increasing order. The LSS $P_{LS}(\delta_E)$ is then defined as the distribution of energy differences between consecutive energy levels
\beq
\delta_{E_j}=E_{S,j+1}-E_{S,j}
\eeq 
within sector $S$.

In addition to Fig.~(\ref{fig:ED}) of our main text, we plotted spectral weight functions, density of states and level spacing statistics for more exmaples in Fig.~(\ref{figS:SWmu}) and Fig.~(\ref{figS:moreED}) with various parameter settings. In Fig.~(\ref{figS:SWmu}), we added nonzero chemical potentials. We notice that the effect of chemical potential term is dominantly shifting the energy of the spectral weight of the $\rho$ fermion. In Fig.~(\ref{figS:moreED}), the results display behaviours between that of a free boson theory and a free fermion theory. 

The Hamiltonian for added nonlinear terms in momentum space is
\beq
H_{\text{non}}=\sum_{k=-\infty}^{\infty}\lambda k^3\left[ \left(c'^\dag_\rho(k) c'_\rho(k)-\frac{1}{2}\right) +\left(c'^\dag_n(k) c'_n(k)-\frac{1}{2}\right)\right]. 
\eeq
Together with Eq.~\ref{seq:ed}, our fermions have dispersion relations $\omega_{\rho}(k)=v_{\rho}k+\lambda k^3$ and $\omega_{n}(k)=v_{n}k+\lambda k^3$.

\section{Other Abelian bilayer fermionic FQH systems}\label{App:E}
Although we have focused on the $4/3$ state for detailed analysis of its edge theory, there are many other Abelian bilayer FQH states with two-by-two K-matrix that can be solved in a similar fashion using the method presented in this paper. In this section, we present mathematical patterns of K-matrices of these states. We leave the generalization to larger $K$ matrices to the future studies.

We restrict ourselves in bilayer electronic systems, thus the K-matrix must have odd integers as diagonal entries. The other condition we impose is that the system has one (and only one) leading interaction term which describes electron hopping processes between the two chiral edges. Similar to Eq.~(2) in main text, it assumes a form
\beq
\mathcal{L}_1=-J(x)e^{i\bm{l}\cdot \bm{\phi}}-h.c., \quad J(x)=Je^{ipx}
\eeq where $p$ again originates from the momentum differences between the two edge modes, and we require the scaling dimension of $e^{i\bm{l}\cdot \bm{\phi}}$ to be two, so that the coupling constant $J$ is again dimensionless and non-negligible at low energies. Besides, this electron hopping process has to be electrically neutral, namely, charge conserving. 

To be more explicit, for a generic two-by-two matrix $K=\begin{pmatrix} a& b\\ b & c\end{pmatrix}$ with boson fields basis $\bm{\phi}=(\phi_1,\phi_2)$, we must have $a,c \in 2\mathbb{Z}_{>0}-1$, $b\in \mathbb{Z}_{\geq 0}$ and $\det K=ac-b^2>0$ (as ground state degeneracy must be positive). The above conditions for the interaction term can be summarized as two equations for the two-component integer vector $\bm{l}=(l_1,l_2)$:
\beq
\frac{1}{2}\bm{l}^T K^{-1}\bm{l}=\frac{a l_2^2-2bl_1l_2+c l_1^2}{2\det K}=2,~\quad \bm{l}^T K^{-1}\bm{q}=\frac{(c-b)l_1+(a-b)l_2}{\det K}=0\ ,
\label{seq:K}
\eeq where the first equation fixes its scaling dimension to be two and the second equation ensures the interaction term is electrically neutral.

Putting on all these constraints, by an extensive search of solutions of Eq.~(\ref{seq:K}), we find two classes of K-matrices, one for filling balanced bilayer systems, and the other for the filling imbalanced case. 

For the balanced case, for each $n\in 2\mathbb{Z}_{>0}$, at filling factor $\nu=\frac{4}{4+n^2}$, we have a series of K-matrix
\beq
\{K^n_j=\begin{pmatrix} \frac{n^2}{2}+j & j\\ j & \frac{n^2}{2}+j\end{pmatrix}\}_{j\in \mathbb{Z}_{>0}}\eeq that satisfies the above characterizations with integer vector $\bm{l}=\left(\pm n,\mp n\right)^T$ describing the marginal interaction term. We see that the $(3,3,1)$-state \cite{Halperin1983,Halperin1984} is the first member of this series. 

The imbalanced case is more complicated, we find three different series of K-matrices. The first series occur at filling factor $\nu=\frac{4}{8n-5}$ for $n\in \mathbb{Z}_{>0}$. There are two varieties,  
\beq
\{K^n_j=\begin{pmatrix} 2n+1 & 2n+3j+1\\ 2n+3j+1 & 2n+4j^2+6j+1\end{pmatrix}\}_{j\in \mathbb{Z}_{>0}}
\eeq
with integer vector $\bm{l}=(\pm 3, \pm (4j+3))^T$ and 
\beq
\{K^{n}_{j \text{ odd}}=\begin{pmatrix} 2n-1 & \frac{4n-j-3}{2}\\ \frac{4n-j-3}{2} & 2n+j^2+j-1\end{pmatrix}\}_{j\leq 4n-3 }, \quad \{K^n_{j \text{ even}}=\begin{pmatrix} 2n-1 & \frac{4n+j-2}{2}\\ \frac{4n+j-2}{2} & 2n+j^2+j-1\end{pmatrix}\}_{j>0}
\eeq
with integer vector $\bm{l}=(\pm 1, \mp (2j+1))^T$ and $\bm{l}=(\pm 1, \pm (2j+1))^T$ respectively. The $(1,7,2)$-state is the second member ($j=2$) of the $n=1$ series.

The second series occurs at filling factor $\nu=\frac{1}{2n}$ for $n\in \mathbb{Z}_{>0}$.
\beq
\{K^n_j=\begin{pmatrix} 2n+1 & 2n+2j+1\\ 2n+2j+1 & 2n+4j^2+4j+1\end{pmatrix}\}_{j\in \mathbb{Z}_{>0}}
\eeq
with integer vector $\bm{l}=(\pm 2, \pm (4j+2))^T$.

The third series occurs at filling factor $\nu=\frac{1}{2n-1}$ for $n \in \mathbb{Z}_{>0}$. \beq
\{K^n_j=\begin{pmatrix} 2n+3 & 2n+4j+3\\ 2n+4j+3 & 2n+4j^2+8j+3\end{pmatrix}\}_{j\in \mathbb{Z}_{>0}}
\eeq
with integer vector $\bm{l}=(\pm 4, \pm (4j+4))^T$.

We note here that the patterns found in this section are not meant to be complete. Besides, many examples followed from this pattern, taken as bilayer FQH systems, are not observed  experimentally. This is partial due to the fact that their filling factors for each layer is smaller than $\nu=\frac{1}{7}\sim\frac{1}{11}$. In these cases, physically, an energetically more dominant Wigner crystal phase or other competing phases may emerge.
\end{document}